\documentclass{article}

\usepackage{epsfig,amsmath}

\newcommand{\eye}{{\mathrm i}}

\newcommand{\half}{{\tfrac{1}{2}}}
\newcommand{\mhalf}{{-\tfrac{1}{2}}}

\newcommand{\threehalf}{{\tfrac{3}{2}}}
\newcommand{\quart}{{\tfrac{1}{4}}}

\newcommand{\XRA}{\xrightarrow}

\begin{document}

\begin{titlepage}

\hbox{}\vspace{1in}
\begin{center}

\textsc{ \Large
Nuclear Magnetic Resonance Spectroscopy: \\ \smallskip
An Experimentally Accessible Paradigm \smallskip
for Quantum Computing}\footnote{
Expanded version of speech presented to
the Fourth Workshop on Physics and Computation,
Boston University, November 24, 1996.}

\vspace{1in}

\textbf{David G. Cory, Mark D. Price}

Nuclear Engineering, Massachusetts Institute
of Technology, Cambridge, MA 02139, USA

\bigskip

\textbf{and Timothy F. Havel}\footnote{
To whom corresponence should be addressed at
\texttt{havel@euclid.med.harvard.edu}.}

Biological Chemistry and Molecular Pharmacology,
Harvard Medical School, Boston, MA 02115, USA

\vspace{1in}

\emph{Submitted to Physica D, January 17, 1997;
Accepted August 27, 1997}

\bigskip

(PACS codes: 05.30.-d, 75.45.+j, 76.70.Fz, 89.80.+h)


\end{center}

\vfill\eject

\begin{center} \textsc{\large Abstract} \end{center} \medskip

We present experimental results which
demonstrate that nuclear magnetic resonance
spectroscopy is capable of efficiently emulating
many of the capabilities of quantum computers,
including unitary evolution and coherent superpositions,
but without attendant wave-function collapse.
This emulation is made possible by two facts.

The first is that the spin active nuclei in
each molecule of a liquid sample are largely
isolated from the spins in all other molecules,
so that each molecule is effectively
an independent quantum computer.
The second is the existence of a
manifold of statistical spin states,
called \emph{pseudo-pure} states,
whose transformation properties are
identical to those of true pure states.
These facts enable us to operate on coherent
superpositions over the spins in each
molecule using full quantum parallelism,
and to combine the results into deterministic
macroscopic observables via thermodynamic averaging.
We call a device based on these principles an
\emph{ensemble quantum computer}.

Our results show that it is indeed possible
to prepare a pseudo-pure state in a macroscopic
liquid sample under ambient conditions,
to transform it into a coherent superposition,
to apply elementary quantum logic gates to this superposition,
and to convert it into the equivalent of an entangled state.
Specifically, we have:
\begin{itemize}
\item Implemented the quantum XOR gate in two different ways,
one using Pound-Overhauser double resonance, and the other
using a spin-coherence double resonance pulse sequence.
\item Demonstrated that the square root of the
Pound-Overhauser XOR corresponds to a conditional
rotation, thus obtaining a universal set of gates.
\item Devised a spin-coherence implementation
of the Toffoli gate, and confirmed that it transforms
the equilibrium state of a four-spin system as expected.
\item Used standard gradient-pulse techniques in NMR to
equalize all but one of the populations in a two-spin system,
so obtaining the pseudo-pure state that corresponds to $|00\rangle$.
\item Validated that one can identify which basic pseudo-pure
state is present by transforming it into one-spin superpositions,
whose associated spectra jointly characterize the state.
\item Applied the spin-coherence XOR gate to a one-spin superposition
to create an entangled state, and confirmed its existence by detecting
the associated double-quantum coherence via gradient-echo methods.
\end{itemize}

\end{titlepage}

\markboth{CORY, PRICE \& HAVEL}%
{NMR: A PARADIGM FOR QUANTUM COMPUTING}

\section{Introduction}
The theory of quantum computing is advancing at a
rate that vastly outstrips its experimental realization
(for accounts, see \cite{Bennett:95,Brassard:95,DiVincenzo:95a}).
Most attempts to implement a quantum computer have
utilized submicroscopic assemblies of quantum spins,
which are difficult to prepare, isolate, manipulate and observe.
A ``homologous'' system that exhibits many of the same properties,
but is easier to work with, would clearly be very useful
both as a means of testing the theoretical predictions,
and exploring implementation issues like error correction.
Such a system is provided by weakly polarized macroscopic
ensembles of spins, which are readily manipulated and observed
by {\em nuclear magnetic resonance spectroscopy\/}, or NMR.

The spins of a molecule in solution are largely isolated from
their surroundings by simple surface-to-volume considerations,
and from the spins in neighboring molecules by diffusional motion,
which averages their dipole-dipole coupling
to a second-order effect \cite{Slichter:90}.
This fact enables us to work with a
{\em reduced density matrix\/} $\pmb \Psi$ of size $2^n$,
where $n$ is the number of spin $\half$ nuclei in the molecule,
rather than $2^N$ where $N$ is the total number
of such spins in the sample \cite{Goldman:88}.
It is also customary in NMR spectroscopy to shift the
reduced density matrix by subtraction of its mean trace,
since only the traceless part undergoes unitary evolution,
and to scale it to have integral elements \cite{ErnBodWok:87}.
In the next paragraph, we define a manifold of
statistical spin states with a reduced density matrix
whose traceless part is proportional to the traceless
part of the usual density matrix of a pure state.

Henceforth, whenever we use the term ``density matrix'', we mean
``reduced, shifted and scaled density matrix'' unless otherwise stated.
When such a density matrix has rank equal to one
(after adding an appropriate multiple of the unit matrix to it),
it can be factored into a dyadic product of the coordinates of
a ``spinor'' and its conjugate versus the usual $\mathbf I_z$ basis,
and this factorization is unique up to an overall phase factor.
This mapping between spinor coordinates and density matrices that can
be shifted to a signature of $[\pm 1,0,\ldots,0]$ is \emph{covariant},
in the sense that if we apply a unitary matrix to the spinor's coordinates,
the corresponding density matrix transforms by
conjugation with the {\em same\/} unitary matrix.
As a result, we can regard such a
density matrix as a kind of spinor,
and perform essentially arbitrary unitary
transformations on it via NMR spectroscopy,
thereby ``emulating'' a quantum computer.
We shall call the states described by density matrices
with $2^n-1$ equal eigenvalues ``pseudo-pure'' states,
and the corresponding spinors ``pseudo-spinors''.

Of course, some things are lost in translation.
For example, the density matrix is not changed on
rotation by $2\pi$, although spinors change sign.
Since these sign changes cannot easily be observed,
this seems to be of little consequence for quantum computing.
More important is the fact that the ``coherence''
observed by NMR spectroscopy is always an ensemble average
over an astronomical number of microscopic quantum systems.
As a consequence, the NMR spectrum of a pseudo-pure state
yields the expectation values of certain observables
relative to the corresponding pseudo-spinor,
rather than a random eigenvalue of one of them.
In particular, \emph{wave function collapse does not occur\/}.
A variety of other more easily controlled ``filtering''
mechanisms are available in NMR spectroscopy, however,
and we have shown that for most computational purposes
the ability to measure expectation values directly is
actually a great advantage \cite{CorFahHav:97}.
NMR experiments on liquid samples possess a
number of other highly desirable features as well;
in particular, the decoherence times are
typically on the order of seconds.

NMR spectroscopy in fact provides a means
of building a nonconventional computer that
can be programmed much like a quantum computer,
but is much easier to implement on at least a limited scale.
In some respects, this approach also resembles DNA computing,
in that it can use the parallelism inherent in
ensembles of molecules to efficiently count the
number of solutions to combinatorial problems,
trading an exponential growth in the time required
against an exponential growth in the sample size.
More generally, we have called a computational
device that operates by running a large number of
quantum computers on coherent superpositions,
and then estimates the expectation values of
observables by summing them over all the quantum
computers, an \emph{ensemble quantum computer}.
A detailed introduction to the theory of such
machines may be found in \cite{CorFahHav:97};
this paper will describe how basic quantum logic
gates can be implemented via NMR spectroscopy, and
present experimental results to validate our claims.
After the majority of these results
had been obtained \cite{CorFahHav:96},
we learned of a similar approach proposed by
other researchers \cite{GershChuan:97,GerChuLlo:96}.

\section{Basic techniques from NMR}
This section introduces the basic techniques from
NMR spectroscopy that are needed for this paper,
and in the process defines the notation
it uses (for more complete introductions,
see e.g. \cite{Canet:96,FarraHarri:95,MateeValer:93}).

Let us consider the simplest nontrivial
case, which contains all the essential
ingredients of solution NMR spectroscopy.
This is a liquid consisting of identical molecules
each containing exactly two coupled spin $\half$ nuclei
of the same isotope (throughout this paper, ${}^1$H).
The dipolar coupling between the spins is averaged to zero
by the rotational motion of the molecules in the liquid,
and hence the coupling in this case is
the so-called \emph{scalar} coupling,
which is mediated by electron correlation
in the chemical bonds linking the atoms.
It will simplify our presentation if we assume
\emph{weak coupling}, i.e.~that the coupling constant $J_{12}$
is small compared to the difference $|\omega_1-\omega_2|$
in the resonance frequencies of the two spins.
With the convention that the magnetic field is along the z-axis,
the Hamiltonian of this system is ${\bf H} = \omega_1 \mathbf I_z^1
+ \omega_2 \mathbf I_z^2 + 2 \pi J_{12} \, \mathbf I_z^1 \mathbf I_z^2$,
where $\mathbf I_z^k$ ($k = 1,2$) are the usual matrices
for the z-component of the angular momentum of the spins.
Because the energy level differences are
small compared to $kT$ at room temperature,
the equilibrium density matrix of this system
is given to an excellent approximation by
$\mathbf{Exp}(-\mathbf H/kT) ~\approx~ \mathbf 1 - \mathbf H/kT$.
On taking account of the fact that $|J_{12}| \ll \omega_1 \approx \omega_2\,$,
removing the trace and scaling, the equilibrium density matrix becomes
\begin{equation}
\pmb \Psi_{eq} = \mathbf I_z^1 + \mathbf I_z^2 =~
\begin{pmatrix}
~1~ & ~0~ & ~0~ & ~0~ \\ ~0~ & ~0~ & ~0~ & ~0~ \\
~0~ & ~0~ & ~0~ & ~0~ \\ ~0~ & ~0~ & ~0~ & -1 \end{pmatrix}
\quad \begin{matrix}
\downarrow\downarrow \\ \downarrow\uparrow \\
\uparrow\downarrow \\ \uparrow\uparrow \end{matrix}
\label{eqn:equil2}
\end{equation}
(where the longitudinal spin states which label the rows and
columns of this matrix are shown along its right-hand side).

Rather than writing them out explicitly,
NMR spectroscopists typically represent their density
matrices as linear combinations of ``product operators'',
i.e.\ products of the usual angular momentum operators
$\mathbf I_x^k$, $\mathbf I_y^k$, $\mathbf I_z^k$
(as in Eq.~(\ref{eqn:equil2})) \cite{ErnBodWok:87,MateeValer:93}.
This makes it very easy to describe the unitary transformations
effected by applying RF (radio-frequency) pulses to the sample.
For example, a ``soft'' pulse whose frequency range
spans the resonance frequency of only the first spin,
and which imparts an energy sufficient to rotate that
spin by an angle $\varphi$ about the $y$-axis
(in a frame rotating with the carrier of
the receiver \cite{ErnBodWok:87}) is
\begin{equation}
\mathbf I_z^1 \XRA{[\varphi]_y^1\;}
\cos(\varphi) \mathbf I_z^1 + \sin(\varphi) \mathbf I_x^1 ~.
\end{equation}
Thus when a ``soft'' (i.e.\ spin-selective) $[\pi/2]$ $y$-pulse
is applied to the first spin of a two-spin system at equilibrium,
we obtain $\mathbf I_x^1 + \mathbf I_z^2$,
while a ``hard'' (nonselective) $[\pi/2]$ $y$-pulse
yields $\mathbf I_x^1 + \mathbf I_x^2$.

The density matrix evolves according to the
time-dependent unitary transformation \cite{ErnBodWok:87}
\begin{equation}
\pmb\Psi \XRA{t\,\mathbf H\;}
\mathbf{Exp}(-\eye\,t\,\mathbf H)\,\pmb\Psi\,
\mathbf{Exp}(\eye\,t\,\mathbf H) ~.
\end{equation}
Since all three terms of the Hamiltonian commute,
the above propagator factors into a product of
the chemical shift and scalar coupling propagators.
The chemical shift propagator for the first spin
can be expanded as
\begin{equation}
\mathbf{Exp}(\eye \,t\,\omega_1\mathbf I_z^1) ~=~
\cos(\omega_1 t/2) \mathbf 1 + 2 \eye\, \sin(\omega_1 t/2) \,\mathbf I_z^1 ~.
\end{equation}
An exercise in the Pauli matrix algebra then shows that
\begin{equation}
\mathbf{Exp}(-\eye \,t\,\omega_1\mathbf I_z^1) \; \mathbf I_x^1 \;
\mathbf{Exp}( \eye \,t\,\omega_1\mathbf I_z^1) ~=~
\cos(\omega_1 t) \mathbf I_x^1 + \sin(\omega_1 t) \,\mathbf I_y^1 ~.
\end{equation}
Altogether, we obtain:
\begin{equation} \begin{aligned} 
\mathbf I_x^1 \XRA{\omega_1 t\, \mathbf I_z^1\;} & ~
\cos(\omega_1 t) \,\mathbf I_x^1 + \sin(\omega_1 t) \,\mathbf I_y^1 \\
\mathbf I_y^1 \XRA{\omega_1 t\, \mathbf I_z^1\;} & ~
\cos(\omega_1 t) \,\mathbf I_y^1 - \sin(\omega_1 t) \,\mathbf I_x^1 \\
\mathbf I_z^1 \XRA{\omega_1 t\, \mathbf I_z^1\;} & ~ \mathbf I_z^1
\end{aligned} \end{equation}
This propagator has no effect on terms involving only the second spin,
which evolve analogously under their own chemical shift Hamiltonian.

The scalar coupling propagator, on the other hand, is
\begin{equation}
\mathbf{Exp}(2\pi \, \eye \, J_{12} \,t\, \mathbf I_z^1 \mathbf I_z^2)
~=~ \cos(\pi J_{12} t/2) \mathbf 1 + 4 \eye\, \sin(\pi J_{12} t/2) \,
\mathbf I_z^1 \mathbf I_z^2 ~.
\end{equation}
A similar calculation shows that it transforms
the one-spin operators as follows:
\begin{equation} \begin{aligned}
\mathbf I_x^1 \XRA{2\pi J_{12} t\, \mathbf I_z^1 \mathbf I_z^2\;} & ~
\cos(\pi J_{12} t) \,\mathbf I_x^1 + 2\, \sin(\pi J_{12} t) \,
\mathbf I_y^1 \mathbf I_z^2 \\
\mathbf I_y^1 \XRA{2\pi J_{12} t\, \mathbf I_z^1 \mathbf I_z^2\;} & ~
\cos(\pi J_{12} t) \,\mathbf I_y^1 - 2\, \sin(\pi J_{12} t) \,
\mathbf I_x^1 \mathbf I_z^2 \\
\mathbf I_z^1 \XRA{2\pi J_{12} t\, \mathbf I_z^1 \mathbf I_z^2\;} & ~
\mathbf I_z^1
\end{aligned} \end{equation}
with analogous expressions for the terms
$\mathbf I_x^2$, $\mathbf I_y^2$ and $\mathbf I_z^2$.

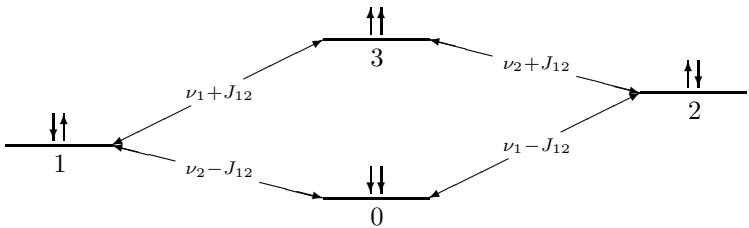
\begin{figure}[bt] \begin{center}
\begin{picture}(300,100)
\put(148,32){\vector(0,-1){10}} \put(152,32){\vector(0,-1){10}}
\put(148,10){$0$}
\put(28,52){\vector(0,-1){10}} \put(32,42){\vector(0,1){10}}
\put(28,30){$1$}
\put(268,62){\vector(0,1){10}} \put(272,72){\vector(0,-1){10}}
\put(268,50){$2$}
\put(148,82){\vector(0,1){10}} \put(152,82){\vector(0,1){10}}
\put(148,70){$3$}
\thicklines
\put(130,20){\line(1,0){40}}
\put(10,40){\line(1,0){40}}
\put(250,60){\line(1,0){40}}
\put(130,80){\line(1,0){40}}
\thinlines
\put(74,34){\vector(-4,1){24}} \put(106,26){\vector(4,-1){24}} \put(78,29)
{$\scriptstyle{\nu_{\scriptscriptstyle 2}-J_{\scriptscriptstyle{12}}}$}
\put(200,35){\vector(-2,-1){30}} \put(220,45){\vector(2,1){30}} \put(198,38)
{$\scriptstyle{\nu_{\scriptscriptstyle 1}-J_{\scriptscriptstyle{12}}}$}
\put(80,55){\vector(-2,-1){30}} \put(100,65){\vector(2,1){30}} \put(78,58)
{$\scriptstyle{\nu_{\scriptscriptstyle 1}+J_{\scriptscriptstyle{12}}}$}
\put(194,74){\vector(-4,1){24}} \put(226,66){\vector(4,-1){24}} \put(198,69)
{$\scriptstyle{\nu_{\scriptscriptstyle 2}+J_{\scriptscriptstyle{12}}}$}
\end{picture}
\caption{ The energy level diagram for a coupled two-spin system,
where the resonance frequencies of the two spins are $\nu_1$
and $\nu_2$, while the coupling constant is $J_{12}$.
The single-quantum transitions allowed by the selection rules
for angular momentum are indicated by two-headed arrows,
together with the energy associated with each. }
\label{fig:transitions}
\end{center} \end{figure}

Physically, the above expressions describe the precession
of the transverse magnetization about the applied field,
which generates a rotating magnetic moment in the $xy$-plane.
The complex-valued signal induced in the receiver
is calculated by taking the trace of the product
of this time-dependent density matrix with
\begin{equation}
\mathbf I_+^1 + \mathbf I_+^2 = \begin{pmatrix}
~0~ & ~1~ & ~1~ & ~0~ \\ ~0~ & ~0~ & ~0~ & ~1~ \\
~0~ & ~0~ & ~0~ & ~1~ \\ ~0~ & ~0~ & ~0~ & ~0~
\end{pmatrix} ~,
\end{equation}
where $\mathbf I_+^k = \mathbf I_x^k + \eye\,
\mathbf I_y^k$ as usual (see e.g.\ \cite{MateeValer:93}).
Since this matrix contains only four nonzero elements,
the spectrum contains direct information on only four
of the ten independent elements of the density matrix,
namely $\Psi_{12} = \Psi_{21}^*, \Psi_{13} = \Psi_{31}^*,
\Psi_{24} = \Psi_{42}^*$ and $\Psi_{34} = \Psi_{43}^*$.
These elements are called \emph{single-quantum coherences},
because they connect pairs of states related by
single spin flips (cf.\ Fig.~\ref{fig:transitions}).

\begin{figure}[bt]
\begin{picture}(100,100)(15,0) \put(10,0){
\scalebox{0.65}[0.30]{\psfig{file=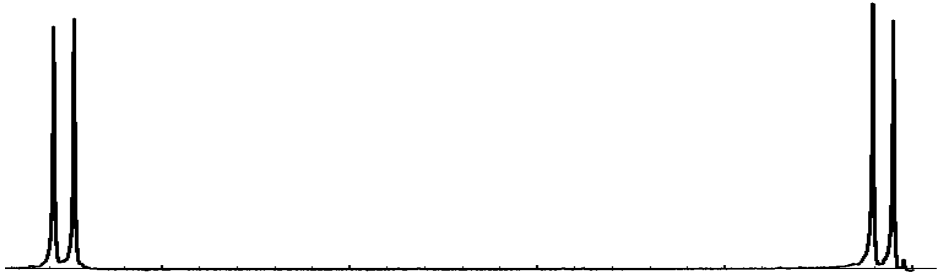}}
} \end{picture}
\caption{
The experimental NMR spectrum of liquid
$2,3$-dibromo-thiophene (see below) in a $9.4$ Tesla field,
at which the two doublets are separated by $130$ Hz.
This spectrum was collected by applying a nonselective
$[\pi/2]$ $y$-pulse to the equilibrium state and
Fourier transforming the resulting signal (see text).
Note that frequency (the horizontal axis)
increases from right-to-left in NMR spectra.}
\label{fig:eq2_hard}
\end{figure}

\begin{figure}[hbt]
\begin{picture}(200,110)(15,0)
\put(110,10){ \scalebox{0.3}[0.3]{\psfig{file=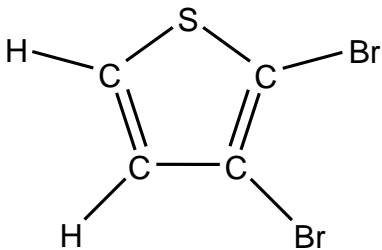}}
} \end{picture} \caption{
The chemical structure of 2,3-dibromo-thiophene.
This molecule is a liquid at room temperature,
and contains two inequivalent, spin $\half$
hydrogen atoms with a coupling constant of $6.0$ Hz. }
\label{fig:two_spin_mol}
\end{figure}

For example, the signal due to the chemical
shift precession of $\mathbf I_x^1$ alone is
\begin{equation}
\mathrm{tr}\bigl(\mathbf I_+^1\, (\cos(\omega_1 t) \mathbf I_x^1
+ \sin(\omega_1 t) \mathbf I_y^1 )\bigr) ~=~
\cos(\omega_1 t) + \eye\, \sin(\omega_1 t) ~.
\end{equation}
These two terms are modulated by the scalar coupling evolution to
\begin{equation} \begin{aligned}
\, & \cos(\omega_1 t)\, \mathrm{tr}\bigl(\mathbf I_+^1\,
( \cos(\pi J_{12} t) \mathbf I_x^1 + 2\, \sin(\pi J_{12} t)\,
\mathbf I_y^1 \mathbf I_z^2 ) \bigr) ~, \\
& \quad =~ \cos(\omega_1 t) \cos(\pi J_{12} t) \\
& \sin(\omega_1 t)\, \mathrm{tr}\bigl(\mathbf I_+^1\,
( \cos(\pi J_{12} t) \mathbf I_x^1 - 2\, \sin(\pi J_{12} t)\,
\mathbf I_x^1 \mathbf I_z^2 ) \bigr) \\
& \quad =~ \sin(\omega_1 t) \cos(\pi J_{12} t) ~.
\end{aligned} \end{equation}
The total signal due to the first spin is thus
\begin{equation} \begin{aligned}
\, & \cos(\omega_1 t) \cos(\pi J_{12} t) +
\sin(\omega_1 t) \cos(\pi J_{12} t) \\
=~ & \half \bigl( \exp(\eye (\omega_1 - \pi J_{12}) t)
+ \exp(\eye (\omega_1 + \pi J_{12}) t) \bigr) ~,
\end{aligned} \end{equation}
which shows that the real part of the Fourier transform
consists of a pair of peaks centered on the resonance
frequency of the spin and separated by the coupling constant;
this is called an \emph{in-phase} doublet,
meaning that both peaks have the same sign.
Thus if the magnetization due to both spins is rotated
into the transverse plane by a hard $[\pi/2]$ $y$-pulse,
one obtains a spectrum containing a pair of doublets.
This is shown in Fig.~\ref{fig:eq2_hard}, using the molecule
$2,3$-dibromo-thiophene shown in Fig.~\ref{fig:two_spin_mol}.

Once the magnetization due to a spin
has been placed in the transverse plane,
a soft $[\pi]$ pulse in the middle of an evolution
period $t$ may be used to \emph{refocus} its chemical
shift evolution during that period, as follows:
\begin{equation} \begin{aligned}
\mathbf I_x^1 \XRA[\quad\quad\quad]{\omega_1 t\,\mathbf I_z^1 / 2} & ~
\cos(\omega_1 t/2) \mathbf I_x^1 + \sin(\omega_1 t/2) \mathbf I_y^1 \\
\XRA[\quad\quad\quad]{ [\pi]_y^1 } & ~
-\cos(\omega_1 t/2) \mathbf I_x^1 + \sin(\omega_1 t/2) \mathbf I_y^1 \\
\XRA[\quad\quad\quad]{\omega_1 t\,\mathbf I_z^1 / 2} & ~
- \cos(\omega_1 t/2) \bigl( \cos(\omega_1 t/2) \mathbf I_x^1
+ \sin(\omega_1 t/2) \mathbf I_y^1 \bigr) \\ & ~
+ \sin(\omega_1 t/2) \bigl( \cos(\omega_1 t/2) \mathbf I_y^1
- \sin(\omega_1 t/2) \mathbf I_x^1 \bigr)
\end{aligned} \end{equation}
Standard trigonometric identities show that
this last expression is simply $-\mathbf I_x^1$.
Thus, up to an inconsequential overall phase factor,
we have cancelled the effect of chemical shift evolution.

A very similar calculation shows that such a soft $[\pi]$
$y$-pulse also cancels the scalar coupling evolution,
so that only the chemical shift evolution of
the other spin occurs during the interval, i.e.
\begin{equation} \begin{aligned}
\, & ~ \mathbf I_x^1 + \mathbf I_x^2 \\
\XRA{t\, \mathbf H /2 - [\pi]_y^1 - t\, \mathbf H /2 - [\pi]_y^1 \;} & ~
\mathbf I_x^1 + \cos(\omega_2 t) \mathbf I_x^2
+ \sin(\omega_2 t) \mathbf I_y^2 ~.
\end{aligned} \end{equation}
We shall denote such a rotation of the second
spin about the $z$-axis by $[\omega_2 t]_z^2$.
Note, however, that $[\pi]_z^2$ is more
easily obtained as $[\pi]_y^2 - [\pi]_x^2$
(a $[\pi]_y^2$ pulse immediately followed by a $[\pi]_x^2$).

A hard $[\pi]$ $y$-pulse applied in the middle of a period,
on the other hand, cancels the chemical shift evolution of
both spins while allowing their scalar coupling to evolve.
This is because the pulse just changes the sign of
the density matrix half-way through the period, i.e.
\begin{equation} \begin{aligned}
\mathbf I_x^1
\XRA[\quad\quad\quad\quad]{\pi J_{12} t\,\mathbf I_z^1 \mathbf I_z^2} & ~
\cos(\pi J_{12} t/2) \mathbf I_x^1 + 2\, \sin(\pi J_{12} t/2) \,
\mathbf I_y^1 \mathbf I_z^2 \\
\XRA[\quad\quad\quad\quad]{[\pi]_y^{12}} & ~ 
-\cos(\pi J_{12} t/2) \mathbf I_x^1 - 2\, \sin(\pi J_{12} t/2) \,
\mathbf I_y^1 \mathbf I_z^2 \\
\XRA[\quad\quad\quad\quad]{\pi J_{12} t\,\mathbf I_z^1 \mathbf I_z^2} & ~
-\cos(\pi J_{12} t) \mathbf I_x^1 - 2\, \sin(\pi J_{12} t) \,
\mathbf I_y^1 \mathbf I_z^2 ~,
\end{aligned} \end{equation}
thereby yielding the negative of what one would
have obtained with no $[\pi]$ pulse (as shown).
We will denote such a scalar coupling evolution,
with no chemical shift evolution, by $[t]$.

A \emph{gradient pulse} produces a transient field
inhomogeneity along the $z$-axis, which has the Hamiltonian
\begin{equation}
\mathbf H_{\mathrm{grad}} ~=~ \gamma \, \mathbf r \cdot \nabla B_z
\end{equation}
where $\mathbf r$ is the position vector within the sample.
Although the microscopic evolution remains unitary (of course!),
a gradient pulse dephases the macroscopic transverse magnetization
due to the off-diagonal elements of the density matrix.
The decay rate of each element is proportional
to the difference in the number of ``up''
spins between the corresponding pair of states,
which is commonly called the \emph{coherence order}.
The net effect is to zero the off-diagonal elements of
the density matrix whose coherence order is nonzero,
thus effecting a \emph{projection} of the state.

This makes it possible to destroy
(or more precisely, render unobservable)
the magnetization due to selected spins
with a combination of soft and gradient pulses, e.g.
\begin{equation}
\mathbf I_z^1 + \mathbf I_z^2 ~ \XRA{[\pi/2]_y^2\;} ~
\mathbf I_z^1 + \mathbf I_x^2 ~ \XRA{[\mathrm{grad}]_z\;} ~
\mathbf I_z^1 ~.
\end{equation}
A further $[\pi/2]$ $y$-pulse transforms
the remaining term to $\mathbf I_x^1$,
which is converted by a scalar coupling evolution of
$[1/(2J_{12})]$ to $2\, \mathbf I_y^1 \mathbf I_z^2$,
and thereafter evolves as
\begin{equation} \begin{aligned}
\cos(\omega_1 t) \, & \bigl(
2\, \cos(\pi J_{12} t) \, \mathbf I_y^1 \mathbf I_z^2
- \sin(\pi J_{12} t) \, \mathbf I_x^1 \bigr) \\
- \sin(\omega_1 t) \, & \bigl(
2\, \cos(\pi J_{12} t) \, \mathbf I_x^1 \mathbf I_z^2 
+ \sin(\pi J_{12} t) \, \mathbf I_x^1 \bigr)
\end{aligned} \end{equation}
Taking the trace product of this with
$\mathbf I_+^1 + \mathbf I_+^2$ yields the signal
\begin{equation} \begin{aligned}
\, & - \cos(\omega_1 t) \sin(\pi J_{12} t)
- \sin(\omega_1 t) \sin(\pi J_{12} t) \\
=~ & \half \bigl( \exp(\eye (\omega_1 - \pi J_{12}) t)
- \exp(\eye (\omega_1 + \pi J_{12}) t) \bigr) ~.
\end{aligned} \end{equation}
Thus if one collects a spectrum
starting after the $[1/(2J_{12})]$ evolution,
one obtains an \emph{anti-phase} doublet
consisting of two peaks of opposite sign.

Alternatively, one can apply a selective $[-\pi/2]_x^2$
pulse to the second spin of $2\, \mathbf I_y^1 \mathbf I_z^2$,
obtaining the correlated state $2\, \mathbf I_y^1 \mathbf I_y^2$,
or a $[\pi/2]_y^2$ pulse to obtain $2\, \mathbf I_y^1 \mathbf I_x^2$.
These contain no single-quantum coherences,
and hence produce no observable magnetization.

\section{The Pound-Overhauser XOR and conditional rotation}
The most obvious way to implement the quantum XOR
(or controlled-NOT) gate is to use a pulse that is
selective for just \emph{one} component of a doublet;
this effects a population transfer similar to the
original ENDOR experiment, and constitutes an example of
\emph{Pound-Overhauser double resonance\/} \cite{Slichter:90}.
Specifically, if we apply a $[\pi]$ pulse about the
$y$-axis to the transitions $1 \leftrightarrow 3$ and
$2 \leftrightarrow 3$ (see Fig.~\ref{fig:transitions}),
we obtain an XOR gate with the output on
the first and second spins, respectively.
These pulses will be denoted by $[\pi]_y^{+k}$ ($k = 1,2$).
If $|\epsilon_1,\epsilon_2\rangle$ denotes the spinor
of a single state ($\epsilon_1,\epsilon_2 \in \{0,1\}$),
these pulses effect the unitary transformations of the
density matrix of the corresponding pure state given by
\begin{equation} \begin{aligned}
\mathbf U_{\mathrm{PO}}^{1\dag}|\epsilon_1,\epsilon_2\rangle
\langle\epsilon_1,\epsilon_2|\mathbf U_{\mathrm{PO}}^1 ~=~ &
| \epsilon_1 \oplus \epsilon_2,\epsilon_2 \rangle
\langle \epsilon_1 \oplus \epsilon_2,\epsilon_2 | \\ &\\
\mathbf U_{\mathrm{PO}}^{2\dag}|\epsilon_1,\epsilon_2\rangle
\langle\epsilon_1,\epsilon_2|\mathbf U_{\mathrm{PO}}^2 ~=~ &
| \epsilon_1, \epsilon_1 \oplus \epsilon_2 \rangle
\langle \epsilon_1, \epsilon_1 \oplus \epsilon_2 | ~,
\end{aligned} \end{equation}
where ``$\oplus$'' denotes the boolean XOR operation, and
\begin{equation} \begin{aligned}
\mathbf U_{\mathrm{PO}}^1 ~\equiv~ &
\begin{pmatrix} ~1~&~0~&~0~&~0~\\ ~0~&~0~&~0~&~1~\\
~0~&~0~&~1~&~0~\\ ~0~&~-1~&~0~&~0~ \end{pmatrix}  \\
\mathbf U_{\mathrm{PO}}^2 ~\equiv~ &
\begin{pmatrix} ~1~&~0~&~0~&~0~\\ ~0~&~1~&~0~&~0~\\ 
~0~&~0~&~0~&~1~\\ ~0~&~0~&~-1~&~0~ \end{pmatrix} ~.
\label{eqn:pox_mat}
\end{aligned} \end{equation}
By applying $[\pi]$ pulses to the transitions $0 \leftrightarrow 1$
and $0 \leftrightarrow 2$, we obtain the boolean operations
$|\epsilon_1 \oplus \bar\epsilon_2, \epsilon_2 \rangle$ and
$|\epsilon_1, \bar\epsilon_1 \oplus \epsilon_2 \rangle$
(where the overbar denotes the NOT of the corresponding qubit);
these pulses will be denoted by $[\pi]_y^{-k}$ ($k = 1,2$).

\newsavebox{\specbase}
\sbox{\specbase}{\line(1,0){50}}
\newsavebox{\specline}
\sbox{\specline}{\line(0,1){10}}

\begin{table}[bt] \begin{center} \begin{minipage}{5in}
\addtolength{\tabcolsep}{-2pt}
\begin{tabular}{|c||c|c||c|c||c|}
\hline
\raisebox{-4pt}{Product} & \multicolumn{2}{c||}{Initial Spectra} &
\multicolumn{2}{c||}{Final Spectra} & \raisebox{-4pt}{Product} \\
\cline{2-5} \raisebox{0pt}[14pt][8pt]{Operator}
& $[\pi/2]_y^1$ & $[\pi/2]_y^2$ & $[\pi/2]_y^1$ & $[\pi/2]_y^2$
& \raisebox{0pt}[14pt][8pt]{Operator} \\
\hline \hline
\raisebox{8pt}{$\mathbf I_z^1$} &
\begin{picture}(50,25)
\put(0,10){\usebox{\specbase}} \put(5,10){\usebox{\specline}}
\put(10,10){\usebox{\specline}}
\end{picture} &
\begin{picture}(50,25)
\put(0,10){\usebox{\specbase}}
\end{picture} &
\begin{picture}(50,25)
\put(0,10){\usebox{\specbase}} \put(5,10){\usebox{\specline}}
\put(10,10){\usebox{\specline}}
\end{picture} &
\begin{picture}(50,25)
\put(0,10){\usebox{\specbase}}
\end{picture} &
\raisebox{8pt}{$\mathbf I_z^1$} \\
\hline
\raisebox{8pt}{$\mathbf I_z^2$} &
\begin{picture}(50,25)
\put(0,10){\usebox{\specbase}}
\end{picture} &
\begin{picture}(50,25)
\put(0,10){\usebox{\specbase}} \put(40,10){\usebox{\specline}}
\put(45,10){\usebox{\specline}}
\end{picture} &
\begin{picture}(50,25)
\put(0,10){\usebox{\specbase}} \put(5,0){\usebox{\specline}}
\put(10,10){\usebox{\specline}}
\end{picture} &
\begin{picture}(50,25)
\put(0,10){\usebox{\specbase}} \put(40,0){\usebox{\specline}}
\put(45,10){\usebox{\specline}}
\end{picture} &
\raisebox{8pt}{$2 \mathbf I_z^1 \mathbf I_z^2$} \\
\hline
\raisebox{8pt}{$2 \mathbf I_z^1 \mathbf I_z^2$} &
\begin{picture}(50,25)
\put(0,10){\usebox{\specbase}} \put(5,0){\usebox{\specline}}
\put(10,10){\usebox{\specline}}
\end{picture} &
\begin{picture}(50,25)
\put(0,10){\usebox{\specbase}} \put(40,0){\usebox{\specline}}
\put(45,10){\usebox{\specline}}
\end{picture} &
\begin{picture}(50,25)
\put(0,10){\usebox{\specbase}}
\end{picture} &
\begin{picture}(50,25)
\put(0,10){\usebox{\specbase}} \put(40,10){\usebox{\specline}}
\put(45,10){\usebox{\specline}}
\end{picture} &
\raisebox{8pt}{$\mathbf I_z^2$} \\
\hline
\end{tabular}
\end{minipage}
\caption{Simplified ``stick'' spectra for each of the three
diagonal product operators that would be observed after
selective $[\pi/2]_y^k$ observation pulses on the $k$-th spin,
before (initial) and after (final) applying an XOR
with its output on the second spin (see text). }
\label{tab:stick_spec}
\end{center} \end{table}
 
\begin{figure}[bt]
\begin{picture}(200,220)(25,0)
\put(20,110){
\scalebox{0.65}[0.30]{\psfig{file=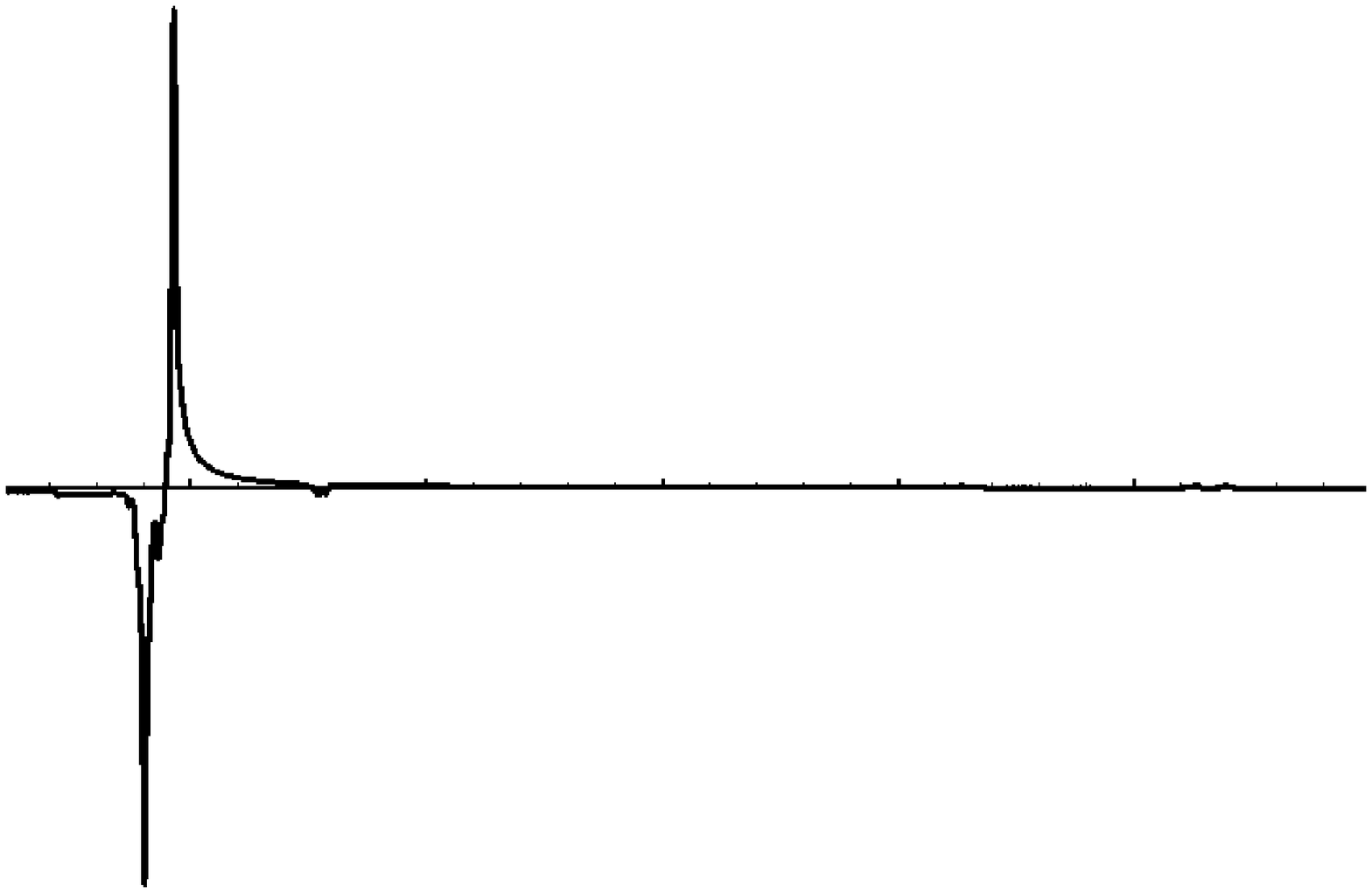}}
} \put(20,0){
\scalebox{0.65}[0.30]{\psfig{file=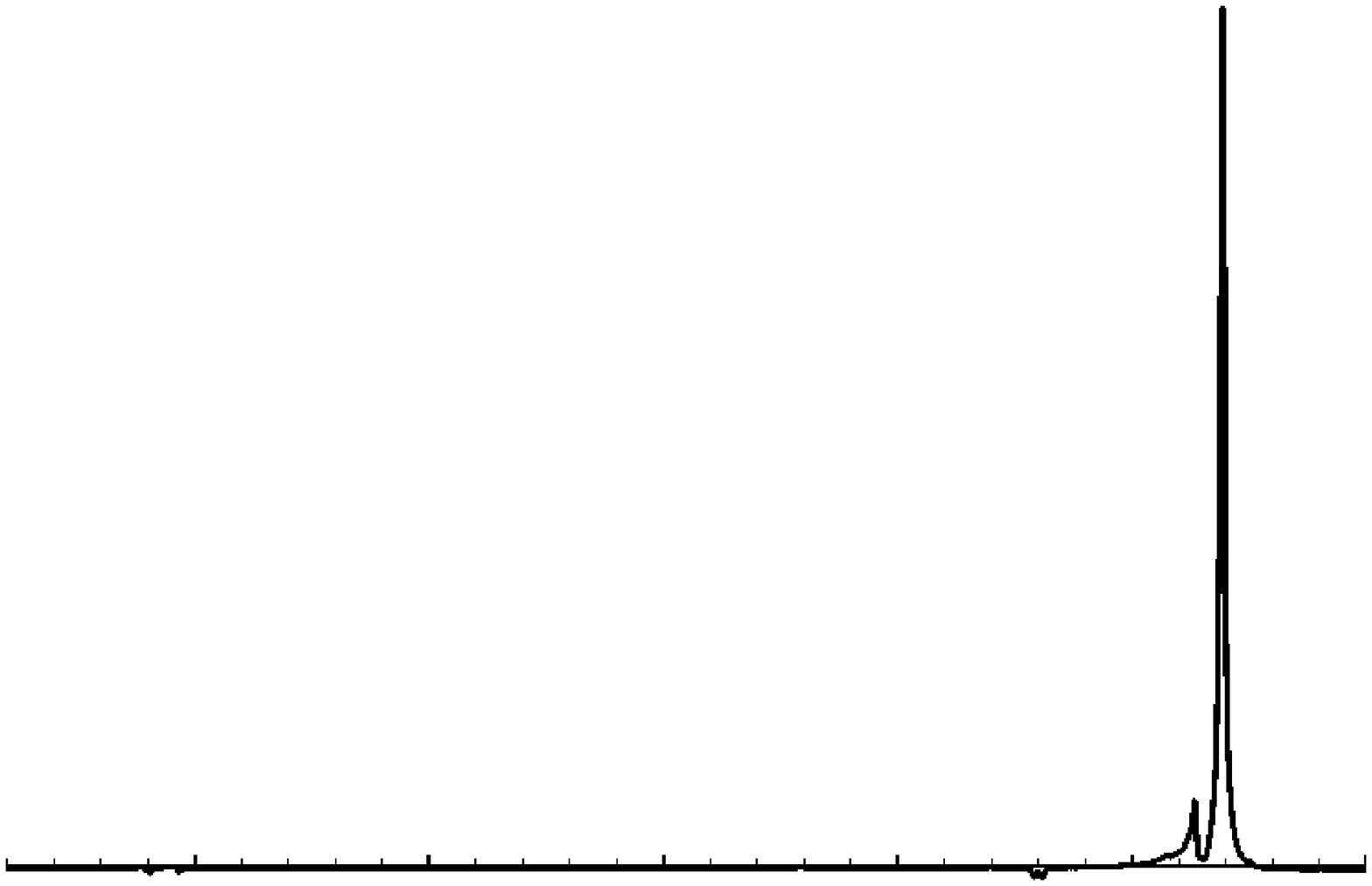}}
} \end{picture}
\caption{ The spectra obtained from $[\pi/2]_y^k$ readout
pulses on the first (above) and second (below) spins,
following a $[\pi]_y^{+1}$ pulse applied
to the equilibrium state (see text). }
\label{fig:eq2_pox1}
\end{figure}

The necessary selectivity can be obtained with a long
sinc-modulated pulse, whose Fourier transform has a square-wave
envelope occupying just the width of a single peak.
Based on the matrices in Eq.~(\ref{eqn:pox_mat}) above,
the result of applying such a pulse to the first
spin (i.e.\ the left-most peak in its doublet)
of a two-spin system at equilibrium should be
\begin{equation} \begin{aligned}
\, & |00\rangle\langle 00| - |11\rangle\langle 11|
= \mathbf I_z^1 + \mathbf I_z^2 \\
\XRA{[\pi]_y^{+1}\;} \quad
& |00\rangle\langle 00| - |01\rangle\langle 01|
= 2\, \mathbf I_z^1 \mathbf I_z^2 + \mathbf I_z^2 ~.
\end{aligned} \end{equation}
This expectation can be confirmed by collecting spectra
following ordinary soft readout pulses on each spin.
The spectra expected from the individual
product operators are shown in a diagrammatic
``stick'' form in Table~\ref{tab:stick_spec},
and the spectrum which results from any sum of
these operators will be the point-by-point sum of
the spectra from the individual terms in the sum.
For example, a readout pulse on the second spin yields
\begin{equation}
2\, \mathbf I_z^1 \mathbf I_z^2 + \mathbf I_z^2
\quad \XRA{[\pi/2]_y^2\;} \quad
2\, \mathbf I_z^1 \mathbf I_x^2 + \mathbf I_x^2
\end{equation}
Due to interference between the in-phase and anti-phase signals,
the resulting spectrum contains only a single peak,
with twice the intensity of the peaks one
gets from a readout on the first spin,
as shown in Fig.~\ref{fig:eq2_pox1}.

\begin{table}[htb] \begin{center}
\renewcommand{\arraystretch}{1.2}
\begin{tabular}{|c||c|c|c|c|} \hline
* & $\half \mathbf 1$ & $\mathbf I_x^2$ &
$\mathbf I_y^2$ & $\mathbf I_z^2$ \\ \hline\hline
$\half \mathbf 1$ & $\half \mathbf 1$ & $-\mathbf I_y^1 \mathbf I_y^2$
& $\mathbf I_y^1 \mathbf I_x^2$ & $\mathbf I_z^2$ \\ \hline
$\mathbf I_x^1$ & $\mathbf I_x^1 \mathbf I_z^2$
& $-\mathbf I_z^1 \mathbf I_x^2$ & $-\mathbf I_z^1 \mathbf I_y^2$
& $\mathbf I_x^1$ \\ \hline
$\mathbf I_y^1$ & $\mathbf I_y^1$ & $-\mathbf I_y^2$
& $\mathbf I_x^2$ & $\mathbf I_y^1 \mathbf I_z^2$ \\ \hline
$\mathbf I_z^1$ & $\mathbf I_z^1 \mathbf I_z^2$
& $\mathbf I_x^1 \mathbf I_x^2$ & $\mathbf I_x^1 \mathbf I_y^2$
& $\mathbf I_z^1$ \\ \hline
\end{tabular}
\caption{ The effect of the $[\pi]_y^{+1}$ pulse
on all the product operators of a two-spin system.
The two factors of the product operator on which the
XOR acts are shown in the left column and top row,
while the result is shown in the corresponding table cell.
Thus for example, a $[\pi]_y^{+1}$ pulse converts the product
operator $\mathbf I_x^1$ to $2\, \mathbf I_x^1 \mathbf I_z^2$. }
\label{tab:pox_prod_ops}
\end{center} \end{table}

Similarly, the result of applying a
Pound-Overhauser XOR to the second spin is
$\mathbf I_z^1 + 2\, \mathbf I_z^1 \mathbf I_z^2$.
The effects of a Pound-Overhauser XOR on all the two-spin
product operators are shown in Table \ref{tab:pox_prod_ops}.
Note that the matrices in eq.~(\ref{eqn:pox_mat})
differ in the sign of one element from the matrices for
the XOR usually encountered in quantum computing, e.g.
\begin{equation}
\mathbf U_{\mathrm{QC}}^1 ~\equiv~
\begin{pmatrix} ~1~&~0~&~0~&~0~\\ ~0~&~0~&~0~&~1~\\ 
~0~&~0~&~1~&~0~\\ ~0~&~1~&~0~&~0~ \end{pmatrix} ~.
\label{eqn:qcx_mat}
\end{equation}
This sign difference has no effect on the results of applying
the gate to the diagonal product operators (i.e.\ single states),
but the results of applying the Pound-Overhauser XOR
to a superposition may differ by phase factors from
the results obtained with the quantum computing XOR.

To see how to get the same phase factors,
we compute the infinitesimal generators of the two XOR's, i.e.
\begin{equation}
\mathbf U_{\mathrm{PO}}^1 ~=~
\mathbf{Exp}(\eye\,\pi\,\pmb\Theta_{\mathrm{PO}}^1)
\quad\text{where}\quad \pmb\Theta_{\mathrm{PO}}^1 ~=~
\frac{1}{2} \begin{pmatrix} ~0~&~0~&~0~&~0~ \\ ~0~&~0~&~0~&-\eye \\
~0~&~0~&~0~&~0~ \\ ~0~&\eye&~0~&~0~ \end{pmatrix}
\end{equation}
and
\begin{equation}
\mathbf U_{\mathrm{QC}}^1 ~=~
\mathbf{Exp}(\eye\,\pi\,\pmb\Theta_{\mathrm{QC}}^1)
\quad\text{where}\quad \pmb\Theta_{\mathrm{QC}}^1 ~=~
\frac{1}{2} \begin{pmatrix} ~0~&~0~&~0~&~0~ \\ ~0~&~1~&~0~&-1 \\
~0~&~0~&~0~&~0~ \\ ~0~&-1&~0~&~1~ \end{pmatrix} ~.
\end{equation}
These generators may be expressed in terms of product operators as
\begin{equation}
\pmb\Theta_{\mathrm{PO}}^1 ~=~ \half \mathbf I_y^1
- \mathbf I_y^1 \mathbf I_z^2
\end{equation}
and
\begin{equation}
\pmb\Theta_{\mathrm{QC}}^1 ~=~ \quart \mathbf 1 - \half \mathbf I_z^2
- \half \mathbf I_x^1 + \mathbf I_x^1 \mathbf I_z^2 ~,
\end{equation}
in both of which all the terms commute with one another.
The term $\quart \mathbf 1$ just leads to an overall phase shift,
while the term $-\half\mathbf I_z^2$ can be
cancelled with a $[\pi/2]_z^2$ rotation.
It follows that the unitary transformation
$\mathbf U_{\mathrm{QC}}^1$ can be obtained
with a $[-\pi]_x^{+1}$ pulse together with
a $[-\pi/2]_z^2$ rotation (in either order).
The effect of this $[XOR]_\mathrm{QC}^1$ pulse
sequence on the two-spin product operators
is shown in Table~\ref{tab:sqx_prod_ops}.
Comparison with Table~\ref{tab:pox_prod_ops} shows
that the two types of XOR operations differ by swaps
of $\half \mathbf 1$ with $\mathbf I_z$ in those
product operators that contain either of these factors,
and $\mathbf I_x$ with $\mathbf I_y$ in all the rest.

\begin{table}[tb] \begin{center}
\renewcommand{\arraystretch}{1.2}
\begin{tabular}{|c||c|c|c|c|} \hline
* & $\half \mathbf 1$ & $\mathbf I_x^2$ &
$\mathbf I_y^2$ & $\mathbf I_z^2$ \\ \hline\hline
$\half \mathbf 1$ & $\half \mathbf 1$ & $\mathbf I_x^1 \mathbf I_x^2$
& $\mathbf I_x^1 \mathbf I_y^2$ & $\mathbf I_z^2$ \\ \hline
$\mathbf I_x^1$ & $\mathbf I_x^1$ & $\mathbf I_x^2$ & $\mathbf I_y^2$
& $\mathbf I_x^1 \mathbf I_z^2$ \\ \hline
$\mathbf I_y^1$ & $\mathbf I_y^1 \mathbf I_z^2$
& $\mathbf I_z^1 \mathbf I_y^2$ & $-\mathbf I_z^1 \mathbf I_x^2$
& $\mathbf I_y^1$ \\ \hline
$\mathbf I_z^1$ & $\mathbf I_z^1 \mathbf I_z^2$
& $-\mathbf I_y^1 \mathbf I_y^2$ & $\mathbf I_y^1 \mathbf I_x^2$
& $\mathbf I_z^1$ \\ \hline
\end{tabular}
\caption{ The effect of the $[XOR]_\mathrm{QC}^1$
pulse sequence on all the product operators of a
two-spin system (cf.\ Table~\ref{tab:pox_prod_ops}). }
\label{tab:sqx_prod_ops}
\end{center} \end{table}

The XOR gate together with arbitrary one-bit rotations,
which one can implement via soft pulses and free precession,
constitute a universal set of quantum logic gates.
It is nevertheless of interest to observe
that the Pound-Overhauser XOR gate generalizes
directly to a conditional rotation.
For example, the square root of the XOR is
obtained by applying a $[\pi/2]_y^{+1}$ pulse
(i.e.\ the same pulse needed to get the XOR,
but for only half as long).
For the output on the first bit, this leads to the matrix
\begin{equation}
\sqrt{\mathbf U_{\mathrm{PO}}^1} ~=~
\frac{1}{\sqrt 2} \begin{pmatrix} ~1~&~0~&~0~&~0~ \\
~0~&~1~&~0~&~1~ \\ ~0~&~0~&~1~&~0~\\ ~0~&-1&~0~&~1~ \end{pmatrix} ~.
\label{eqn:one_bit_mat}
\end{equation}
To get a unitary transformation whose matrix is that
of the square root of the quantum computing XOR, i.e.
\begin{equation}
\sqrt{\mathbf U_{\mathrm{QC}}^1} ~=~ \frac{1+\eye}{\sqrt{2}}
\begin{pmatrix} 1-\eye&~0~&~0~&~0~ \\ ~0~&~1~&~0~&-\eye \\
~0~&~0~&1-\eye&~0~ \\ ~0~&-\eye&~0~&~1~ \end{pmatrix} ~,
\end{equation}
one need only apply a $[-\pi/2]_x^{+1}$ pulse together
with a $[-\pi/4]_z^2$ pulse (in either order).

\begin{figure}[bt]
\begin{picture}(100,220)(25,0)
\put(17,120){
\scalebox{0.65}[0.30]{\psfig{file=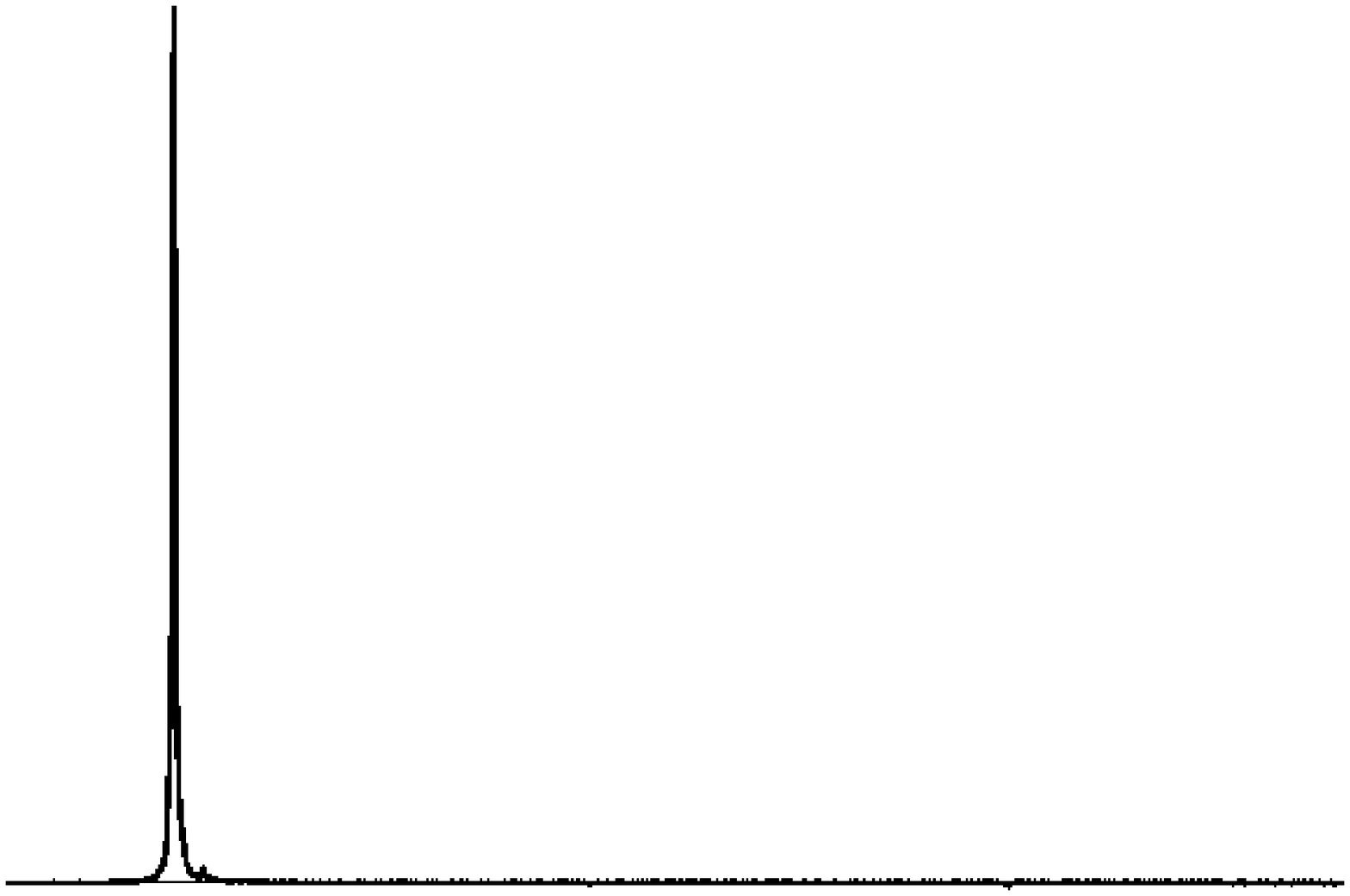}}
} \put(20,0){
\scalebox{0.65}[0.30]{\psfig{file=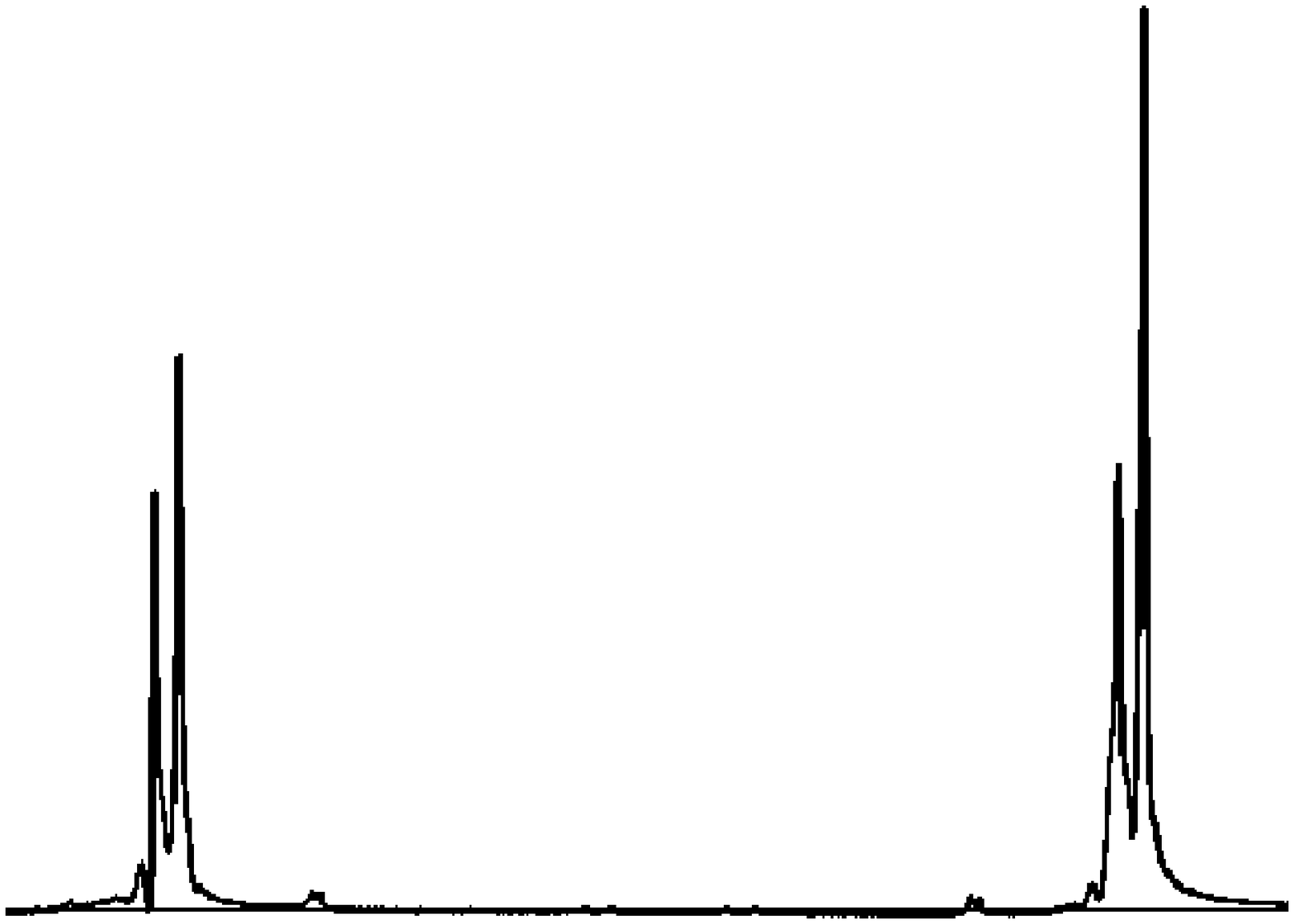}}
} \end{picture}
\caption{ Validation of a Pound-Overhauser conditional rotation.
The upper spectrum was collected immediately after applying
the $[\pi/2]_y^{+1}$ pulse to the equilibrium state,
while the lower spectrum was collected following the
application of an additional hard $[\pi/2]_y^{12}$ pulse. }
\label{fig:one_bit}
\end{figure}

Based on the matrix in Eq.~(\ref{eqn:one_bit_mat}),
the effect of a $[\pi/2]_y^{+1}$ pulse
on the equilibrium state should be:
\begin{equation}
\mathbf I_z^1 + \mathbf I_z^2 ~\XRA{[\pi/2]_y^{+1}\;}~
\half \mathbf I_x^1 - \mathbf I_x^1 \mathbf I_z^2
+ \mathbf I_z^1 \mathbf I_z^2 + \half \mathbf I_z^1 + \mathbf I_z^2
\end{equation}
This gives rise to a spectrum containing a single peak
at the very same frequency that the pulse was tuned to.
An additional hard readout pulse produces the state
\begin{equation}
-\half \mathbf I_z^1 + \mathbf I_z^1 \mathbf I_x^2 +
\mathbf I_x^1 \mathbf I_x^2 + \half \mathbf I_x^1 + \mathbf I_x^2 ~.
\end{equation}
The corresponding spectra are shown in Fig.~\ref{fig:one_bit}.
The general formula for a conditional rotation
in terms of product operators is:
\begin{equation}
\mathbf I_z^1 ~\XRA{[\varphi]_y^{+1}\;}~ \begin{aligned}[t] &
\cos^2(\varphi/2) \mathbf I_z^1 + \cos(\varphi/2) \sin(\varphi/2)
\bigl( \mathbf I_x^1 - 2\, \mathbf I_x^1 \mathbf I_z^2 \bigr) \\
& +\, 2\, \sin^2(\varphi/2) \mathbf I_z^1 \mathbf I_z^2 \end{aligned}
\end{equation}

\begin{table}[tb] \begin{center}
\renewcommand{\arraystretch}{1.2}
\begin{tabular}{|c||c|c|c|c|} \hline
* & $\half \mathbf 1$ & $\mathbf I_x^2$ &
$\mathbf I_y^2$ & $\mathbf I_z^2$ \\ \hline\hline
$\half \mathbf 1$ & $\half \mathbf 1$ & $-\mathbf I_y^1 \mathbf I_y^2$
& $\mathbf I_y^1 \mathbf I_x^2$ & $\mathbf I_z^2$ \\ \hline
$\mathbf I_x^1$ & $\mathbf I_y^1$ & $-\mathbf I_y^2$
& $\mathbf I_x^2$ & $\mathbf I_y^1 \mathbf I_z^2$ \\ \hline
$\mathbf I_y^1$ & $-\mathbf I_x^1 \mathbf I_z^2$
& $\mathbf I_z^1 \mathbf I_x^2$ & $\mathbf I_z^1 \mathbf I_y^2$
& $-\mathbf I_x^1$ \\ \hline
$\mathbf I_z^1$ & $\mathbf I_z^1 \mathbf I_z^2$
& $\mathbf I_x^1 \mathbf I_x^2$ & $\mathbf I_x^1 \mathbf I_y^2$
& $\mathbf I_z^1$ \\ \hline
\end{tabular}
\caption{ The effect of the $[XOR\,]_{\mathrm{SC}}^1$ pulse
sequence on all the product operators of a two-spin system
(cf.\ Tables 2 \& 3). }
\label{tab:scx_prod_ops}
\end{center} \end{table}

\section{The spin-coherence XOR and Toffoli gates}
Due to the degree of selectivity required and
the necessity for direct coupling between the spins involved,
the Pound-Overhauser XOR gate can be difficult to apply.
We have therefore developed the following
pulse sequence, which constitutes an example of
\emph{spin-coherence double resonance\/} \cite{Slichter:90},
and promises to be more generally useful:
\begin{equation}
[XOR]_{\mathrm{SC}}^k ~\equiv~ [\pi/2]_y^k - [1/(2J_{12})]
- [\pi/2]_x^k \quad\quad (k = 1,2)
\end{equation}
To show that this does indeed effect the boolean XOR operation,
we demonstrate that it has the same effect on the diagonal product
operators as the Pound-Overhauser XOR above, e.g.\ for $k = 1$:
\begin{equation} \begin{aligned}
\, & \mathbf I_z^1 \XRA{[\pi/2]_y^1\;} \mathbf I_x^1 \XRA[]{[1/(2J_{12})]\;}
2\, \mathbf I_y^1 \mathbf I_z^2 \XRA{[\pi/2]_x^1\;}
2\, \mathbf I_z^1 \mathbf I_z^2 \\
& \mathbf I_z^2 \XRA{[\pi/2]_y^1\;} \mathbf I_z^2 \XRA[]{[1/(2J_{12})]\;}
\mathbf I_z^2 \XRA{[\pi/2]_x^1\;} \mathbf I_z^2 \\
& 2\, \mathbf I_z^1 \mathbf I_z^2 \XRA{[\pi/2]_y^1\;}
2\, \mathbf I_x^1 \mathbf I_z^2
\XRA{[1/(2J_{12})]\;} \mathbf I_y^1 \XRA[]{[\pi/2]_x^1\;} \mathbf I_z^1
\end{aligned} \end{equation}
Table \ref{tab:scx_prod_ops} shows the effect of
this gate on all the two-spin product operators.
Alternatively, we can just multiply together the
matrices of the three individual steps, to obtain
\begin{equation}
\mathbf U_{\mathrm{SC}}^1 ~=~ \frac{1}{\sqrt{2}}
\begin{pmatrix} 1+\eye &~0~&~0~&~0~\\~0~&~0~&~0~& 1+\eye \\
0 &~0~& 1-\eye &~0~\\~0~& -1+\eye &~0~&~0~\end{pmatrix} ~.
\label{eqn:scx_mat}
\end{equation}
We shall call this the \emph{spin-coherence XOR gate}.

\begin{figure}[hbt]
\begin{picture}(200,220)(25,0)
\put(20,110){
\scalebox{0.65}[0.30]{\psfig{file=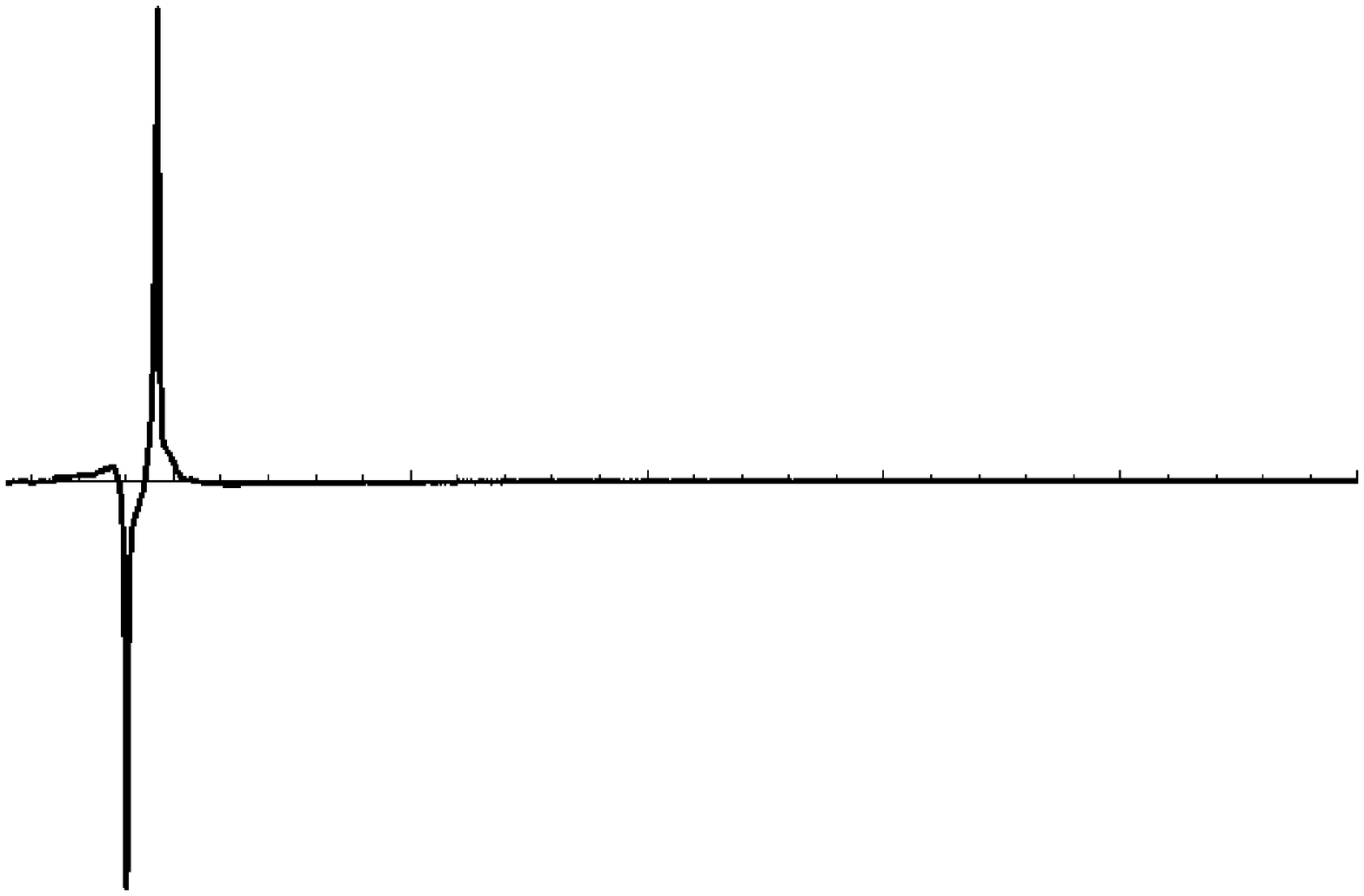}}
} \put(20,0){
\scalebox{0.65}[0.30]{\psfig{file=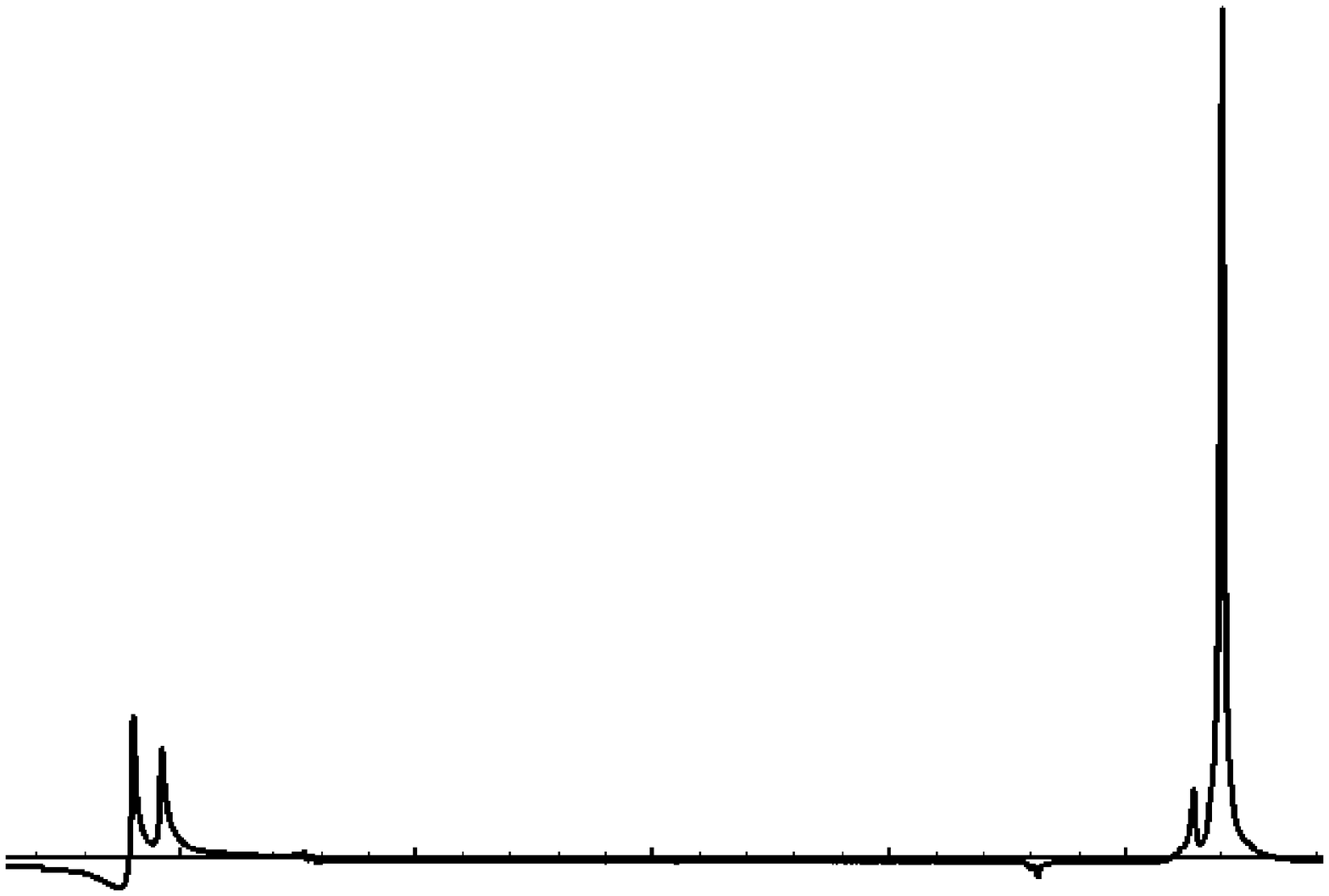}}
} \end{picture}
\caption{ The spectra obtained from readout pulses
on the two spins, following a $[XOR\,]_{\mathrm{SC}}^1$
pulse sequence applied to the equilibrium state (see text). }
\label{fig:eq2_scx1}
\end{figure}

The validity of this XOR implementation is corroborated
by the spectra shown in Fig.~\ref{fig:eq2_scx1},
which were obtained by applying this sequence to
the equilibrium state followed by soft readout pulses.
The infinitesimal generator of the product of the matrix
in Eq.~(\ref{eqn:scx_mat}) with the matrix of the usual
quantum computing XOR from Eq.~(\ref{eqn:qcx_mat}) is
\begin{equation}
\mathbf U_{\mathrm{QC}}^1 \mathbf U_{\mathrm{SC}}^1 ~=~
\mathbf{Exp}(\eye\,\pi\,\pmb\Delta)
\quad\text{where}\quad \pmb\Delta ~=~
\half \mathbf 1 + \half \mathbf I_z^1 - \half \mathbf I_z^2 ~.
\end{equation}
Together with the fact that $\mathbf U_{\mathrm{QC}}^1$ is self-inverse
and commutes with $\mathbf{Exp}(\eye\,\pi\,\mathbf I_z^2/2)$,
this generator shows that the phases of the nonzero components can again
be equalized by composing this pulse sequence with suitable $z$-rotations,
namely $[\pi/2]_z^2 - [XOR]_{\mathrm{SC}}^1 - [-\pi/2]_z^1$.

We have also developed a pulse sequence,
analogous to the above $[XOR\,]_{\mathrm{SC}}^k$
sequence, which transforms the longitudinal spin
states of a three-spin system according to the truth
table of the well-known Toffoli gate \cite{Toffoli:80}.
We call this the \emph{Toffoli pulse sequence}:
\begin{equation} \begin{aligned}
\, [TOF\,]_{\mathrm{SC}}^k ~\equiv~~ &
[\pi/2]_y^k - [1/(4J)] - [\pi/2]_y^k - [1/(4J)] \\
& - [-\pi/2]_x^k - [1/(4J)] - [-\pi/2]_x^k
\end{aligned} \end{equation}
This pulse sequence assumes that the coupling
constants $J_{k\ell}$ and $J_{km}$ have the same value $J$,
which can always be arranged by inverting the
spin whose coupling constant with $k$ is greater
using a soft $[\pi]$ pulse part way through each delay.
This changes the sign of the effective coupling constant,
so that the time-average coupling constant
can be given any desired value between the
original coupling constant and its negative.

\begin{figure}[hbt]
\begin{picture}(100,150)(15,0)
\put(110,10){ \scalebox{0.30}[0.30]{\psfig{file=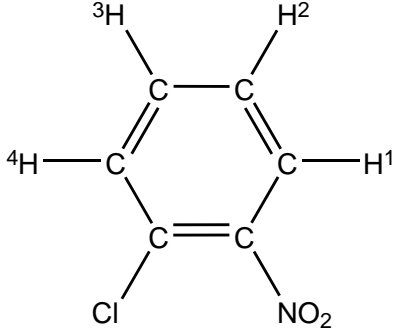}}
} \end{picture} \caption{
The chemical structure of 1-chloro-2-nitro-benzene,
which was used to validate the spin-coherence
implementation of the Toffoli gate by applying the
$[TOF]_{\mathrm{SC}}^1$ sequence to hydrogen atoms $1 - 3$.
The relevant coupling constants among these atoms are $J_{12} = 8.0$ Hz.,
$J_{13} = 1.5$ Hz., and $J_{14} \le 0.1$ Hz., and $J_{23} = 7.0$ Hz. }
\label{fig:four_spin_mol}
\end{figure}

If we place the output on the first spin ($k = 1$),
the matrix of the above Toffoli sequence can be shown to be:
\begin{equation}
\mathbf V_{\mathrm{SC}}^1 ~=~
-\tfrac{1}{\sqrt{2}} \begin{pmatrix}
1+\eye&~0~&~0~&~0~&~0~&~0~&~0~&~0~\\
~0~&~\eye~&~0~&~0~&~0~&~0~&~0~&~0~\\
~0~&~0~&~\eye~&~0~&~0~&~0~&~0~&~0~\\
~0~&~0~&~0~&~0~&~0~&~0~&~0~&1-\eye\\
~0~&~0~&~0~&~0~&1-\eye&~0~&~0~&~0~\\
~0~&~0~&~0~&~0~&~0~&-\eye~&~0~&~0~\\
~0~&~0~&~0~&~0~&~0~&~0~&-\eye~&~0~\\
~0~&~0~&~0~&-1-\eye&~0~&~0~&~0~&~0~
\end{pmatrix}
\label{eqn:tof_mat}
\end{equation}
This is easily seen to act upon the density
matrices associated with single states as
\begin{equation} \begin{aligned}
\, & \mathbf V_{\mathrm{SC}}^{1\dag}|\epsilon_1,\epsilon_2,\epsilon_3\rangle
\langle\epsilon_1,\epsilon_2,\epsilon_3| \mathbf V_{\mathrm{SC}}^1 \\
=~ &
|\epsilon_1\oplus(\epsilon_2\wedge\epsilon_3),\epsilon_2,\epsilon_3\rangle
\langle\epsilon_1\oplus(\epsilon_2\wedge\epsilon_3),\epsilon_2,\epsilon_3|
\end{aligned} \end{equation}
(where the ``$\wedge$'' denotes the boolean AND operation).
Unlike the XOR sequence, however, the Toffoli sequence
does not simply permute the diagonal product operators;
instead, it results in a sum of such diagonal operators, e.g.
\begin{equation} \begin{aligned}
\mathbf I_z^1 ~~
\XRA[\quad]{[\pi/2]_y^1\;} ~~ &
\mathbf I_x^1 \\
\XRA[\quad]{[1/(4J)]\;} ~~ &
\half \mathbf I_x^1 + \mathbf I_y^1 \mathbf I_z^2 + \mathbf
I_y^1 \mathbf I_z^3 - 2\, \mathbf I_x^1 \mathbf I_z^2 \mathbf I_z^3 \\
\XRA[\quad]{[\pi/2]_y^1\;} ~~ &
-\half \mathbf I_z^1 + \mathbf I_y^1 \mathbf I_z^2 + \mathbf
I_y^1 \mathbf I_z^3 + 2\, \mathbf I_z^1 \mathbf I_z^2 \mathbf I_z^3 \\
\XRA[\quad]{[1/(4J)]\;} ~~ &
-\half \mathbf I_z^1 - \half \mathbf I_x^1 - 2 \mathbf I_x^1
\mathbf I_z^2 \mathbf I_z^3 + 2\, \mathbf I_z^1 \mathbf I_z^2 \mathbf I_z^3 \\
\XRA[\quad]{[-\pi/2]_x^1\;} ~~ &
-\half \mathbf I_y^1 - \half \mathbf I_x^1 - 2 \mathbf I_x^1
\mathbf I_z^2 \mathbf I_z^3 + 2\, \mathbf I_y^1 \mathbf I_z^2 \mathbf I_z^3 \\
\XRA[\quad]{[1/(4J)]\;} ~~ &
-\half \mathbf I_y^1 - \mathbf I_y^1 \mathbf I_z^2 - \mathbf I_y^1
\mathbf I_z^3 + 2\, \mathbf I_y^1 \mathbf I_z^2 \mathbf I_z^3 \\
\XRA[\quad]{[-\pi/2]_x^1\;} ~~ &
\half \mathbf I_z^1 + \mathbf I_z^1 \mathbf I_z^2 + \mathbf I_z^1
\mathbf I_z^3 - 2\, \mathbf I_z^1 \mathbf I_z^2 \mathbf I_z^3 ~.
\end{aligned} \end{equation}
Similar calculations lead to the following complete list:
\begin{equation} \begin{aligned}
\mathbf I_z^1 ~~
\XRA{[TOF\,]_{\mathrm{SC}}^1\;} ~~ &
\half \mathbf I_z^1 + \mathbf I_z^1 \mathbf I_z^2 + \mathbf I_z^1
\mathbf I_z^3 - 2 \mathbf I_z^1 \mathbf I_z^2 \mathbf I_z^3 \\
\mathbf I_z^2 ~~
\XRA{[TOF\,]_{\mathrm{SC}}^1\;} ~~ &
\mathbf I_z^2 \\
\mathbf I_z^3 ~~
\XRA{[TOF\,]_{\mathrm{SC}}^1\;} ~~ &
\mathbf I_z^3 \\
2 \mathbf I_z^1 \mathbf I_z^2 ~~
\XRA{[TOF\,]_{\mathrm{SC}}^1\;} ~~ &
\half \mathbf I_z^1 + \mathbf I_z^1 \mathbf I_z^2 - \mathbf
I_z^1 \mathbf I_z^3 + 2 \mathbf I_z^1 \mathbf I_z^2 \mathbf I_z^3 \\
2 \mathbf I_z^1 \mathbf I_z^3 ~~
\XRA{[TOF\,]_{\mathrm{SC}}^1\;} ~~ &
\half \mathbf I_z^1 - \mathbf I_z^1 \mathbf I_z^2 + \mathbf
I_z^1 \mathbf I_z^3 + 2 \mathbf I_z^1 \mathbf I_z^2 \mathbf I_z^3 \\
2 \mathbf I_z^2 \mathbf I_z^3 ~~
\XRA{[TOF\,]_{\mathrm{SC}}^1\;} ~~ &
2 \mathbf I_z^2 \mathbf I_z^3 \\
4 \mathbf I_z^1 \mathbf I_z^2 \mathbf I_z^3 ~~
\XRA{[TOF\,]_{\mathrm{SC}}^1\;} ~~ &
-\half \mathbf I_z^1 + \mathbf I_z^1 \mathbf I_z^2 + \mathbf
I_z^1 \mathbf I_z^3 + 2 \mathbf I_z^1 \mathbf I_z^2 \mathbf I_z^3
\end{aligned} \end{equation}

\begin{figure}[p]
\begin{picture}(400,450)(25,0)
\put(10,300){ \scalebox{0.30}[0.30]{\psfig{file=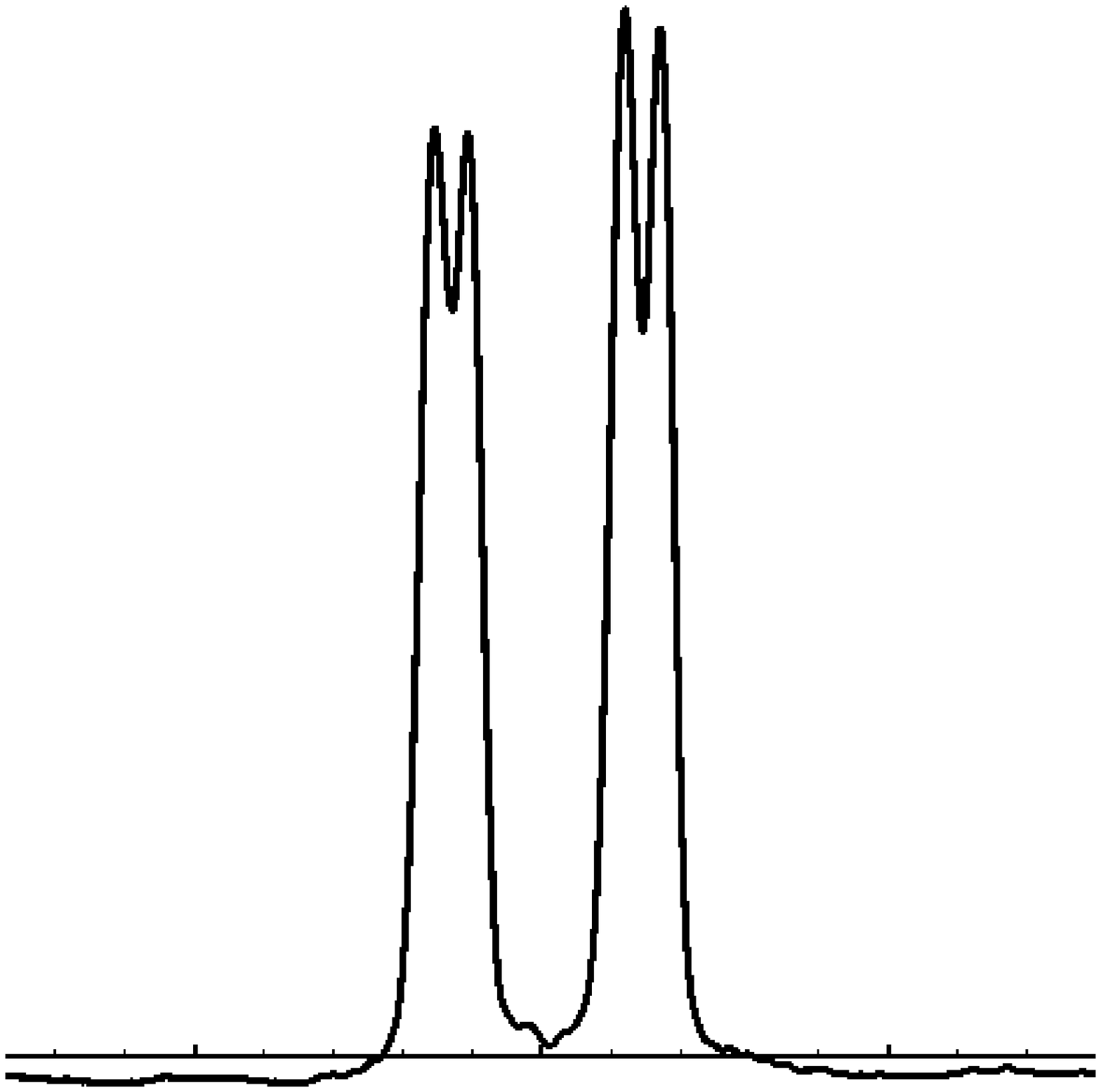}} }
\put(230,300){ \scalebox{0.30}[0.30]{\psfig{file=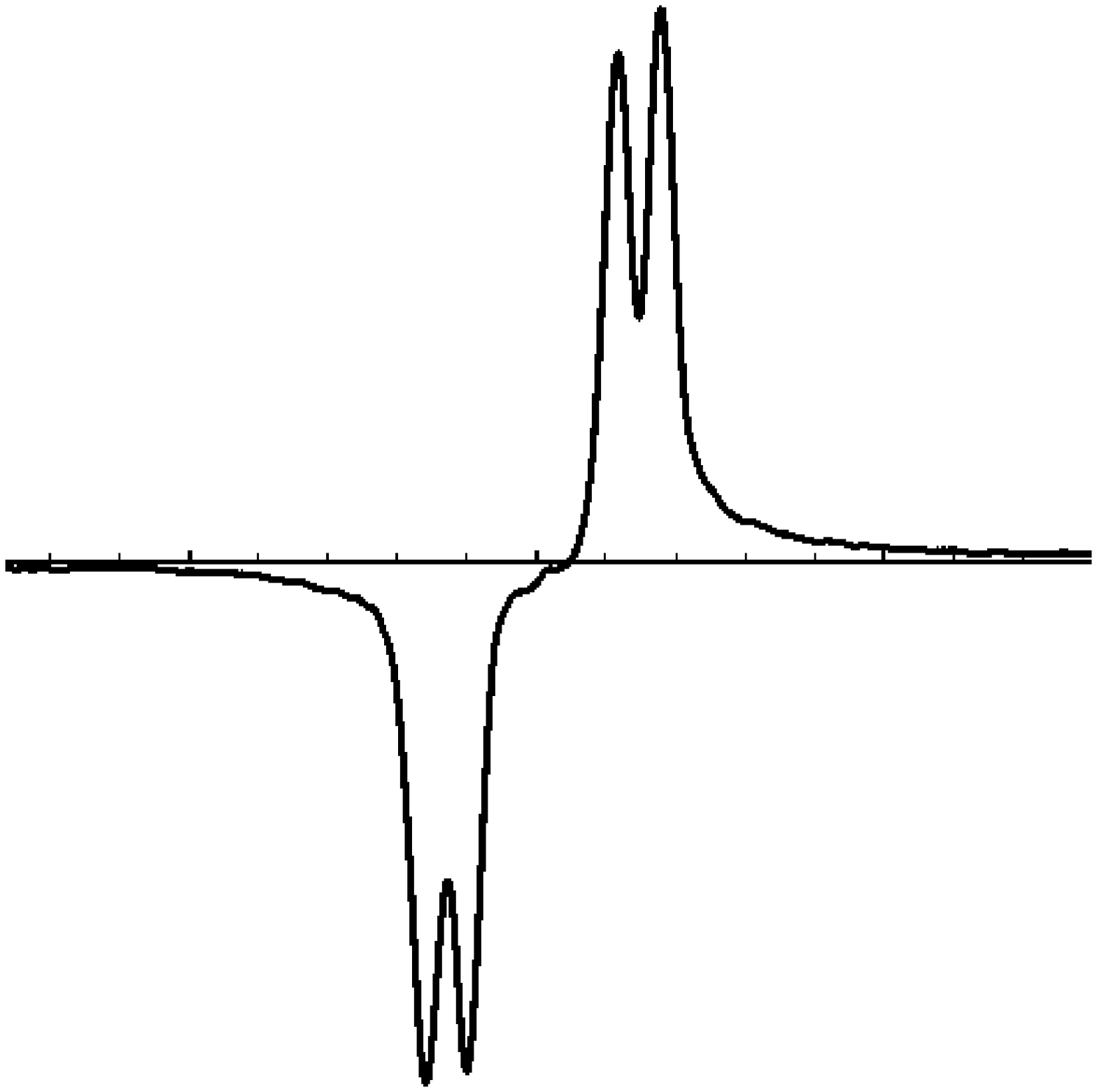}} }
\put(10,150){ \scalebox{0.30}[0.30]{\psfig{file=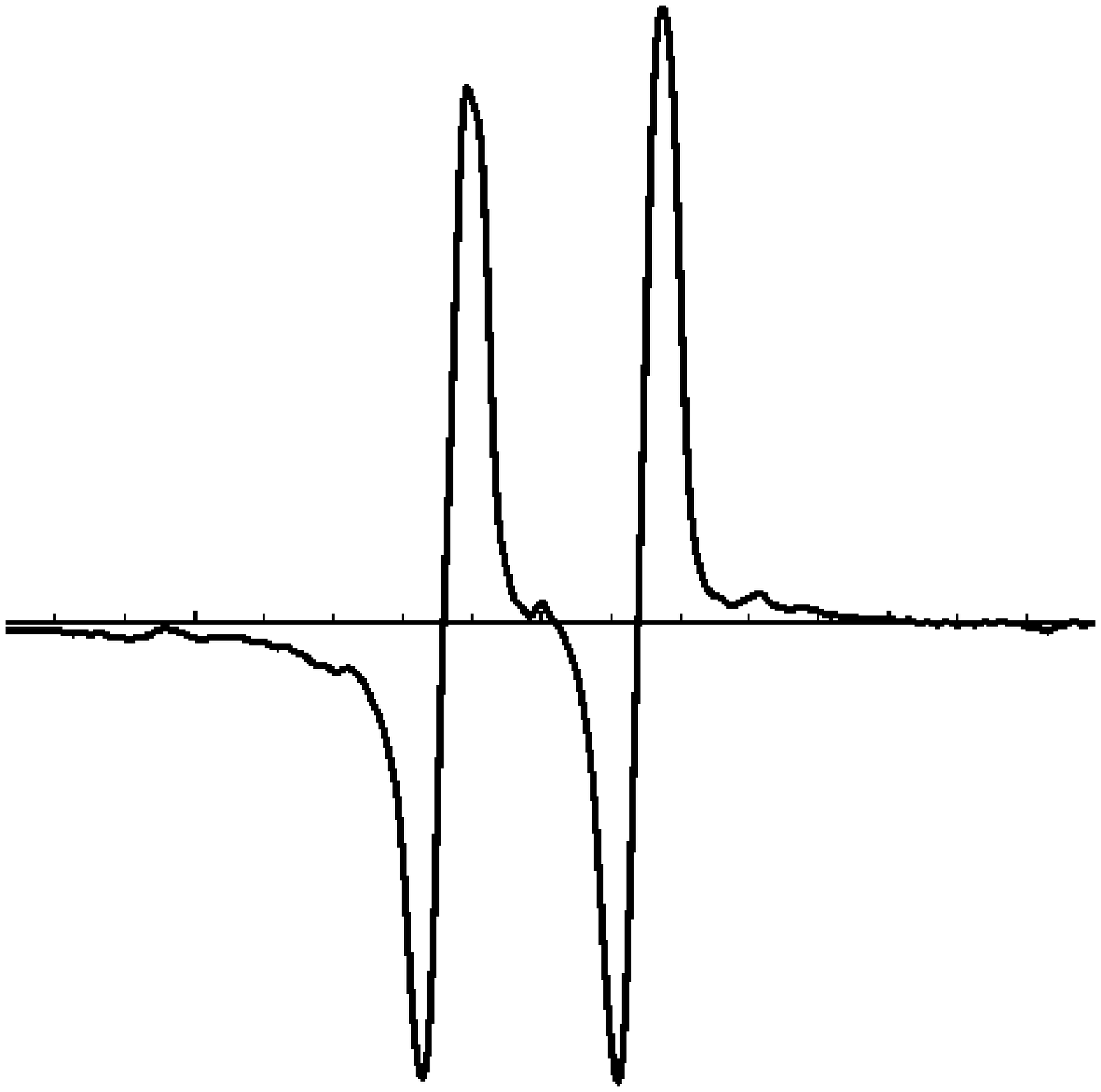}} }
\put(230,150){ \scalebox{0.30}[0.30]{\psfig{file=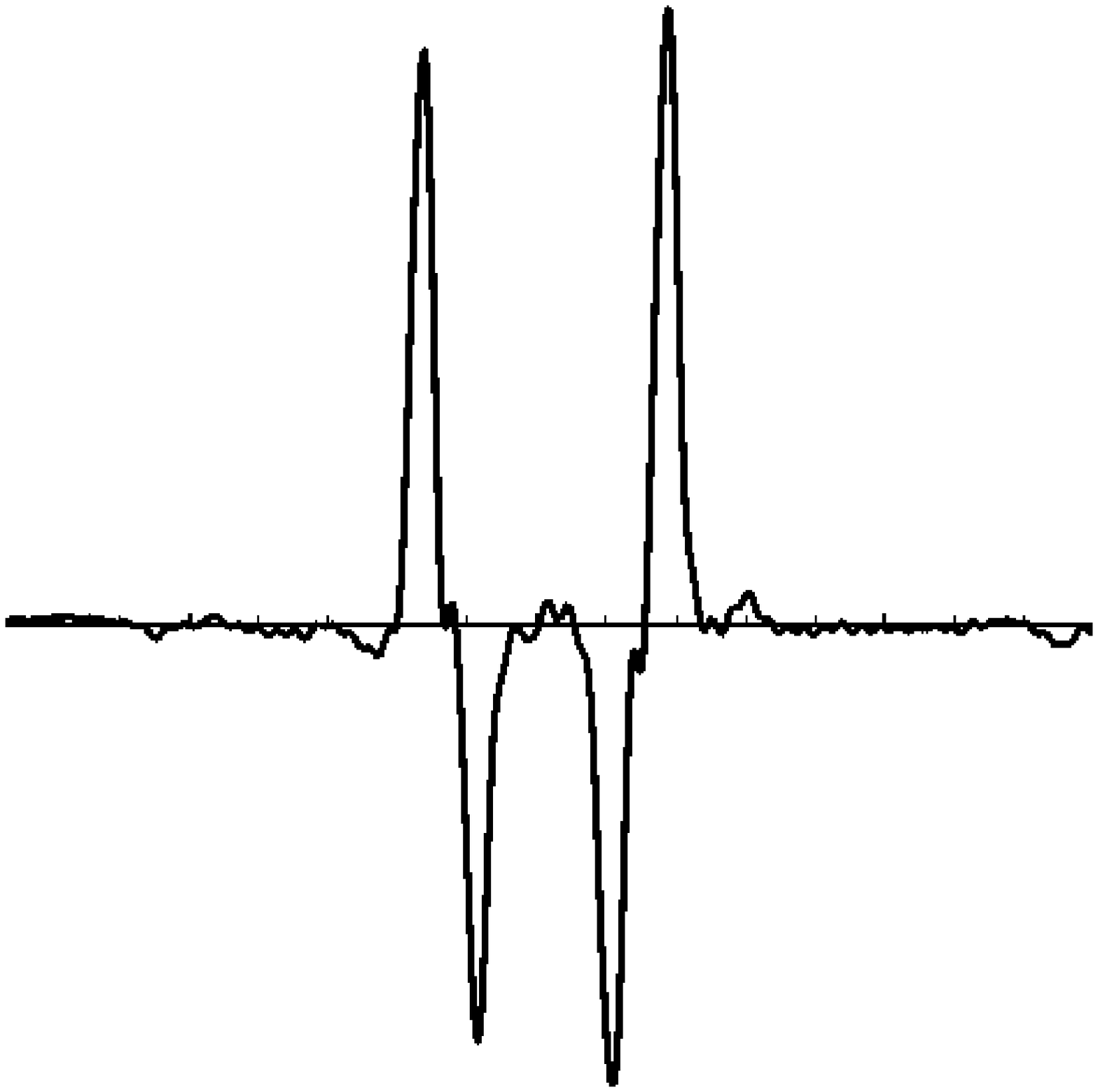}} }
\put(120,0){ \scalebox{0.30}[0.30]{\psfig{file=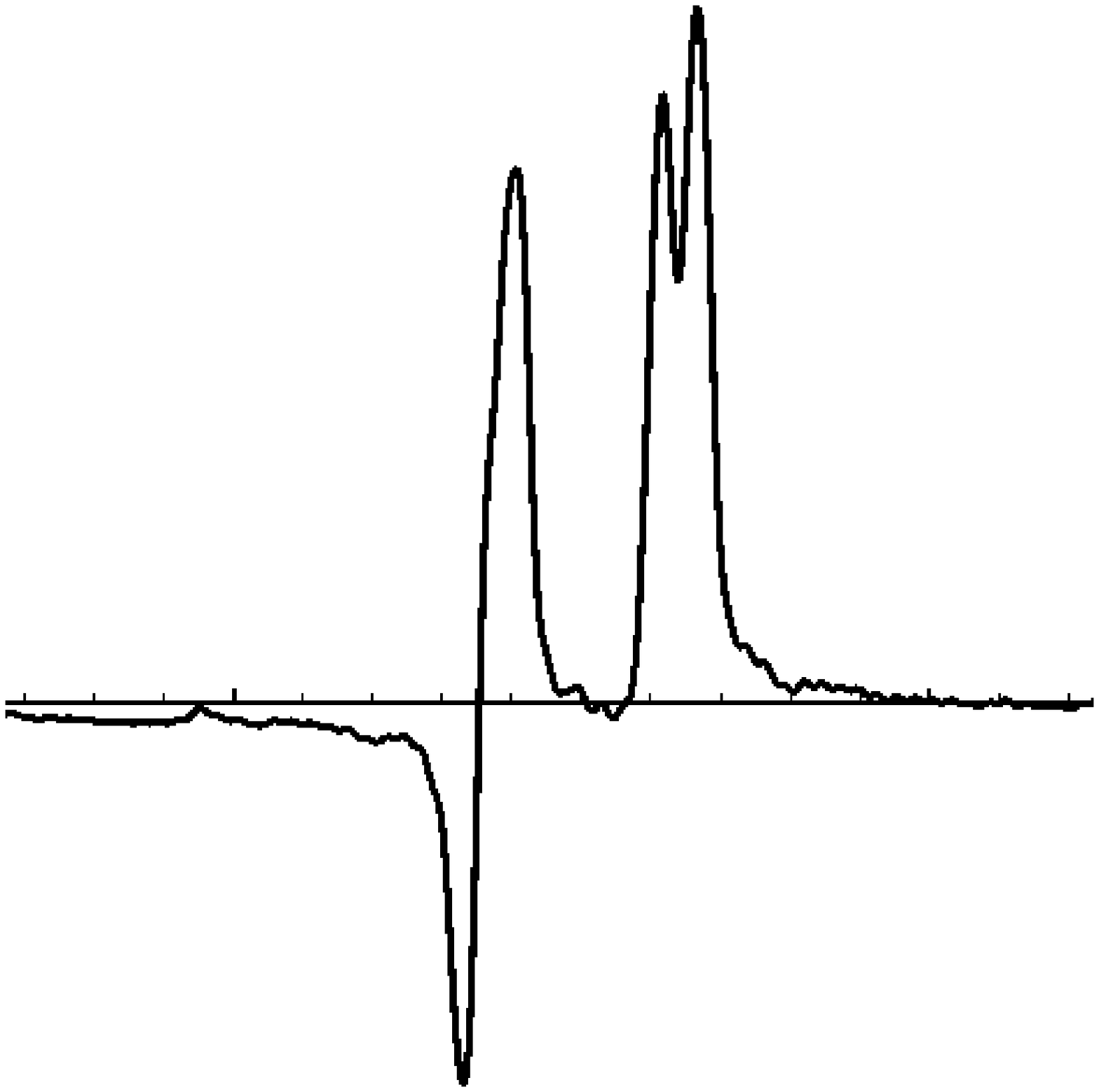}} }
\put(130,340){$\begin{smallmatrix} \mathbf I_x^1
\end{smallmatrix}$}
\put(340,340){$\begin{smallmatrix} 2\, \mathbf I_x^1 \mathbf I_z^2
\end{smallmatrix}$}
\put(120,180){$\begin{smallmatrix} 2\, \mathbf I_x^1 \mathbf I_z^3
\end{smallmatrix}$}
\put(340,180){$\begin{smallmatrix} 4\,\mathbf I_x^1 \mathbf I_z^2
\mathbf I_z^3 \end{smallmatrix}$}
\put(200,20){$\begin{smallmatrix}
\mathbf I_x^1 + 2\,\mathbf I_x^1 \mathbf I_z^2 + \\
2\,\mathbf I_x^1 \mathbf I_z^3 - 4\,\mathbf I_x^1 \mathbf I_z^2
\mathbf I_z^3 \end{smallmatrix}$}
\end{picture}
\caption{ The spectra of 1-chloro-2-nitro-benzene
obtained following a $[\pi/2]_y^1$ readout pulse
to the four diagonal product operators $\mathbf I_z^1$,
$2\,\mathbf I_z^1\mathbf I_z^2$, $2\,\mathbf I_z^1\mathbf I_z^3$
and $2\,\mathbf I_z^1\mathbf I_z^2\mathbf I_z^3$,
as well as to the result of applying the $[TOF]_{\mathrm{SC}}^1$
pulse sequence to the equilibrium state (see text). }
\label{fig:eq4_tof1}
\end{figure}

Once again, these claims can be corroborated by using the matrix
in Eq.~(\ref{eqn:tof_mat}) to predict the result of applying
the $[TOF\,]_{\mathrm{SC}}^1$ sequence to the equilibrium state,
and then verifying that the results are consistent with
the spectra collected after appropriate readout pulses.
These spectra are rather complicated, however,
and hence we shall simply show the result of
applying the readout pulse to the first spin.
The molecule used for these experiments,
1-chloro-2-nitro-benzene, is actually a four-spin system,
thus enabling us to also demonstrate that we can
apply quantum logic gates to subsets of spins,
given sufficient frequency resolution.
This spectrum is shown in Fig.~\ref{fig:eq4_tof1},
along with the spectra obtained by applying the same
readout pulse to the four diagonal product operators
(prepared by the gradient-pulse techniques described earlier)
for comparison.
The fact that just one of the four peaks due to the
first spin has been inverted by the Toffoli sequence
is a direct reflection of the fact that the population
difference of the first spin has been inverted in just
those molecules wherein the other two spins were ``up''.

In order to equalize the phase factors,
we compute the infinitesimal generator of the
product of the matrix $\mathbf V_{\mathrm{SC}}^1$ for the
$[TOF]_{\mathrm{SC}}^1$ gate shown in Eq.~(\ref{eqn:tof_mat})
with the desired matrix $\mathbf V_{\mathrm{QC}}^1$
(consisting of ones at all the same nonzero locations):
\begin{equation} \begin{aligned}
\, & \mathbf{Exp}(\eye\,\pi\,\pmb\Delta) ~=~
\mathbf V_{\mathrm{QC}}^1 \mathbf V_{\mathrm{SC}}^1
\quad\quad\text{where} \\
& \pmb\Delta ~=~ \tfrac{1}{8} \mathbf 1 - \mathbf I_z^1 - \quart
\mathbf I_z^2 - \quart \mathbf I_z^3 - \half \mathbf I_z^1 \mathbf I_z^2
- \half \mathbf I_z^1 \mathbf I_z^3 + \half \mathbf I_z^2 \mathbf I_z^3
\end{aligned} \end{equation}
These are all $z$-rotations of one kind or another,
and hence can all be readily obtained from the three-spin
analogues of the implementations introduced above for
a two-spin system (save possibly for the rotation
whose generator is $\mathbf I_z^2 \mathbf I_z^3$,
which will require a relay sequence if these spins
are not directly coupled \cite{Slichter:90}).

The Toffoli gate is known to be universal for classical computation,
and hence the existence of this pulse sequence shows that any boolean
function can be implemented in NMR via soft pulses selective for single spins.
A Pound-Overhauser implementation of the Toffoli gate is also possible,
but we have not yet done the experiments to demonstrate this.

\section{Superpositions of pseudo-spinors}
If one generates all three of the states $\mathbf I_z^1$,
$\mathbf I_z^2$ and $2\, \mathbf I_z^1 \mathbf I_z^2$
simultaneously in the same sample,
one obtains the density matrix
\begin{equation}
\mathbf I_z^1 + \mathbf I_z^2 + 2\, \mathbf I_z^1 \mathbf I_z^2 =
\begin{pmatrix} \threehalf & ~0~ & ~0~ & ~0~ \\ ~0~ & \mhalf & ~0~ & ~0~ \\
~0~ & ~0~ & \mhalf & ~0~ \\ ~0~ & ~0~ & ~0~ & \mhalf \end{pmatrix}
\label{eqn:pure1}
\end{equation}
This matrix shifts to
\begin{equation}
\begin{pmatrix} 2 & ~0~ & ~0~ & ~0~ \\ ~0~ & ~0~ & ~0~ & ~0~ \\
~0~ & ~0~ & ~0~ & ~0~ \\ ~0~ & ~0~ & ~0~ & ~0~ \end{pmatrix} ~,
\end{equation}
which can in turn be factored into the product
of a pseudo-spinor with its conjugate,
\begin{equation}
2\, | 00 \rangle\langle 00 | ~ \equiv ~ 2 \begin{pmatrix}
1 \\ ~0~ \\ ~0~ \\ ~0~ \end{pmatrix} \begin{pmatrix}
1 & ~0~ & ~0~ & ~0~ \end{pmatrix} ~.
\end{equation}
Therefore, the expression
$\mathbf I_z^1 + \mathbf I_z^2 + 2\, \mathbf I_z^1 \mathbf I_z^2$
in Eq.\ (\ref{eqn:pure1}) represents a pseudo-pure
state in product operator notation.

The following RF and gradient pulse sequence transforms the
equilibrium state of a two-spin system into this pseudo-pure state:
\begin{equation} \begin{aligned}
\, & ~~ \mathbf I_z^1 + \mathbf I_z^2 \\
\XRA[\quad\quad\quad\quad]{[\pi/3]_x^2}
& ~~ \mathbf I_z^1 + \mathbf I_z^2 / 2 - \mathbf I_y^2 \sqrt{3} / 2 \\
\XRA[\quad\quad\quad\quad]{[\text{grad}]_z\;}
& ~~ \mathbf I_z^1 + \mathbf I_z^2 / 2 \\
\XRA[\quad\quad\quad\quad]{[\pi/4]_x^1}
& ~~ \mathbf I_z^1 / \sqrt{2} + \mathbf I_z^2 / 2 - \mathbf I_y^1 / \sqrt{2} \\
\XRA[\quad\quad\quad\quad]{[1/(2J_{12})]\;}
& ~~ \mathbf I_z^1 / \sqrt{2} + \mathbf I_z^2 / 2 +
\sqrt{2}\, \mathbf I_x^1 \mathbf I_z^2 \\
\XRA[\quad\quad\quad\quad]{[-\pi/4]_y^1}
& ~~ \mathbf I_z^1 / 2 + \mathbf I_z^2 / 2 - \mathbf I_x^1 / 2 +
\mathbf I_x^1 \mathbf I_z^2 + \mathbf I_z^1 \mathbf I_z^2 \\
\XRA[\quad\quad\quad\quad]{[\text{grad}]_z\;}
& ~~ \mathbf I_z^1 / 2 + \mathbf I_z^2 / 2 + \mathbf I_z^1 \mathbf I_z^2
\label{eqn:miracle}
\end{aligned} \end{equation}
Subsequently, any of the four \emph{basic} pseudo-pure
states (with a density matrix shifting to a matrix with a
single nonzero element on the diagonal) can be obtained by means
of soft $[\pi]$ pulses (which perform the NOT operation on the spins).
In terms of product operators, these can be written as:
\begin{equation} \begin{aligned}
2\,| 00 \rangle\langle 00 | - \tfrac{1}{2} \mathbf 1 ~ = ~ \quad
& \mathbf I_z^1 + \mathbf I_z^2 + 2\, \mathbf I_z^1 \mathbf I_z^2 \\
2\,| 01 \rangle\langle 01 | - \tfrac{1}{2} \mathbf 1 ~ = ~ \quad
& \mathbf I_z^1 - \mathbf I_z^2 - 2\, \mathbf I_z^1 \mathbf I_z^2 \\
2\,| 10 \rangle\langle 10 | - \tfrac{1}{2} \mathbf 1 ~ = ~ -
& \mathbf I_z^1 + \mathbf I_z^2 - 2\, \mathbf I_z^1 \mathbf I_z^2 \\
2\,| 11 \rangle\langle 11 | - \tfrac{1}{2} \mathbf 1 ~ = ~ -
& \mathbf I_z^1 - \mathbf I_z^2 + 2\, \mathbf I_z^1 \mathbf I_z^2
\label{eqn:basic_pure}
\end{aligned} \end{equation}

The factor of $1/2$ difference between the last line of
Eq.~(\ref{eqn:miracle}) and the first line in Eq.~(\ref{eqn:basic_pure})
is due to a 25\% loss of total polarization during the gradient pulses.
Note that by changing the sign of either of the two $[\pi/4]$
rotations in Eq.~(\ref{eqn:miracle}), we can generate the state
$\mathbf I_z^1 + \mathbf I_z^2 - 2\, \mathbf I_z^1 \mathbf I_z^2$.
This is the \emph{negative} of last state in Eq.~(\ref{eqn:basic_pure}),
which means the polarization of the sample (i.e.\ the sign
of the nonzero element after shifting the density matrix)
is inverted between the two states.
The negatives of all four states in
Eq.~(\ref{eqn:basic_pure}) can likewise
be interconverted by soft $[\pi]$ pulses,
but none of these states can be converted
to its negative by means of RF pulses.
Whether we use the four basis states in
Eq.~(\ref{eqn:basic_pure}) or their negatives
is irrelevant for computational purposes.

\begin{figure}[tb]
\begin{picture}(100,220)(25,0)
\put(20,110){
\scalebox{0.65}[0.30]{\psfig{file=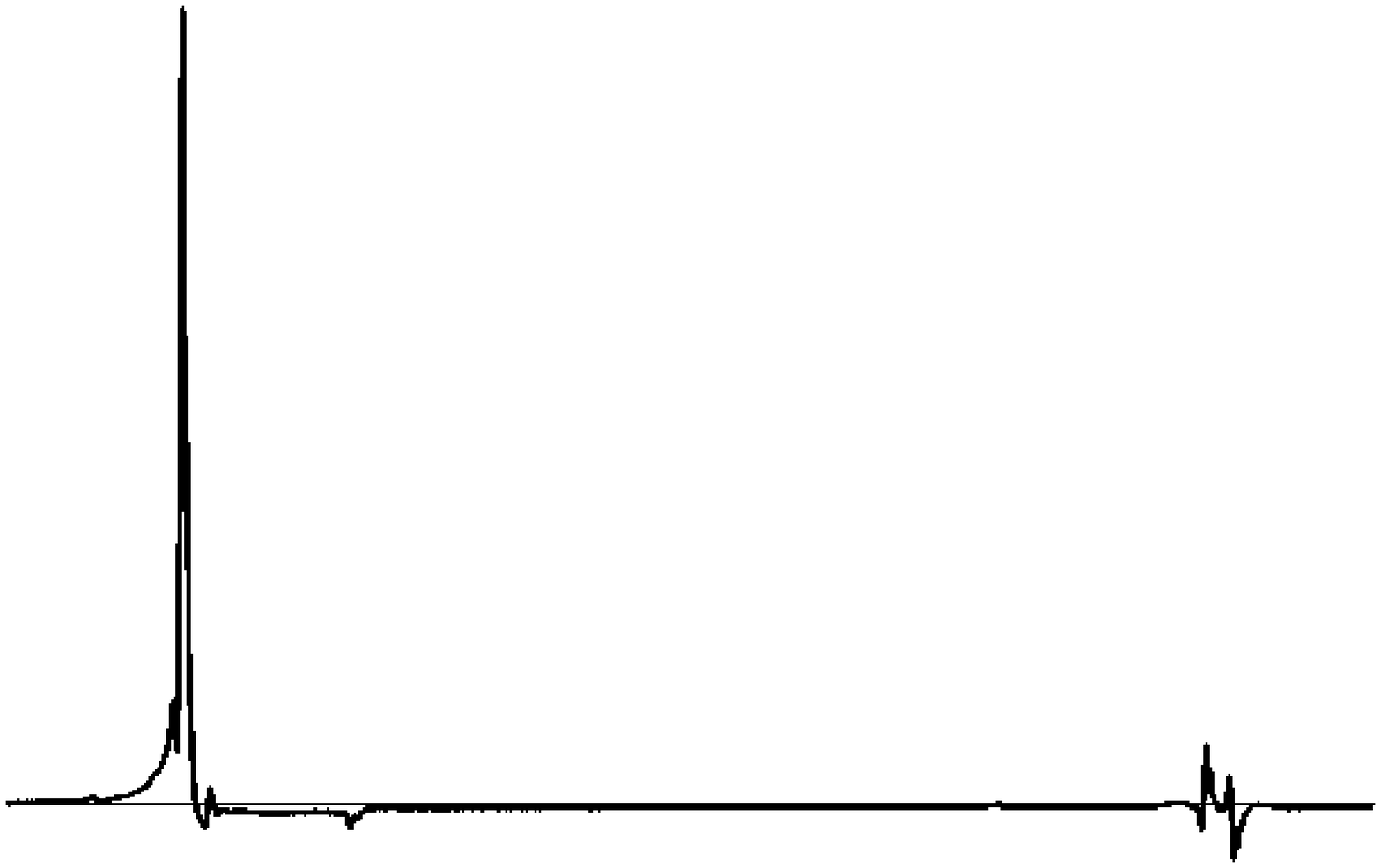}}
} \put(20,0){
\scalebox{0.65}[0.30]{\psfig{file=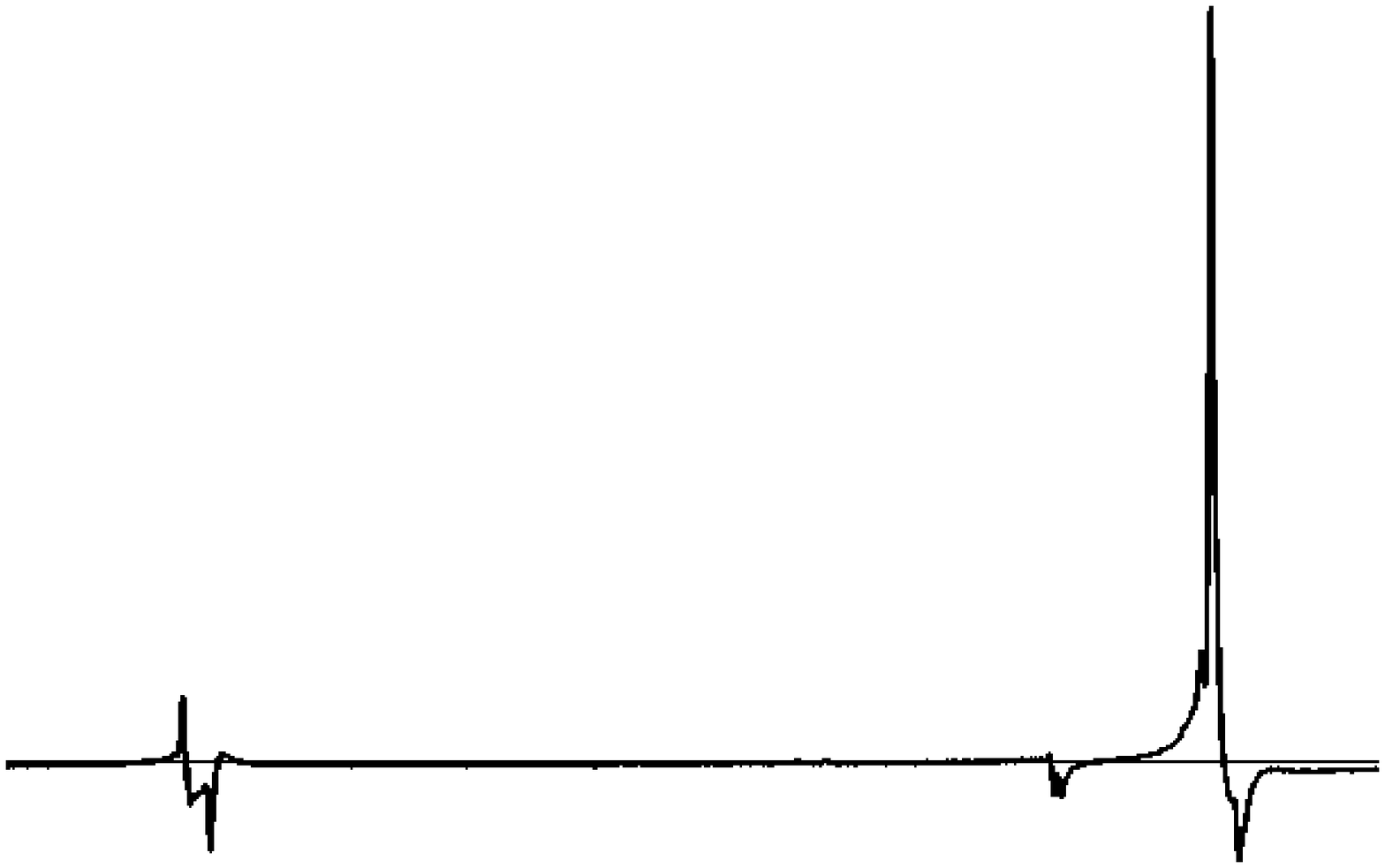}}
} \end{picture}
\caption{
Two experimental NMR spectra of $2,3$-dibromo-thiophene,
which corroborate the creation of the pseudo-pure state
$\mathbf I_z^1 + \mathbf I_z^2 - 2\, \mathbf I_z^1 \mathbf I_z^2$
via readout pulses selective for the first (above)
and second (below) spins (see text). }
\label{fig:pp2_soft}
\end{figure}

To validate that with the use of a $[\pi/4]_y^1$ instead
of a $[-\pi/4]_y^1$ pulse in Eq.~(\ref{eqn:miracle}),
we have actually created the expected state
$\mathbf I_z^1 + \mathbf I_z^2 - 2\, \mathbf I_z^1 \mathbf I_z^2$,
we consider the spectra that are obtained after applying
soft readout pulses to each of the two spins, i.e.
\begin{equation}
\mathbf I_z^1 + \mathbf I_z^2 - 2\, \mathbf I_z^1 \mathbf I_z^2
~\XRA{[\pi/2]_y^1\;}~
\mathbf I_x^1 + \mathbf I_z^2 - 2\, \mathbf I_x^1 \mathbf I_z^2
\end{equation}
and
\begin{equation}
\mathbf I_z^1 + \mathbf I_z^2 - 2\, \mathbf I_z^1 \mathbf I_z^2
~\XRA{[\pi/2]_y^2\;}~
\mathbf I_z^1 + \mathbf I_x^2 - 2\, \mathbf I_z^1 \mathbf I_x^2 ~.
\end{equation}
This creates the difference of an in-phase with an anti-phase
state in both cases, so that the associated spectra consist of
single peaks at the leftmost position of the corresponding doublet.
These spectra are shown in Fig.~\ref{fig:pp2_soft}.

\begin{figure}[htb]
\begin{picture}(100,240)(25,0) \put(20,130){
\scalebox{0.65}[0.30]{\psfig{file=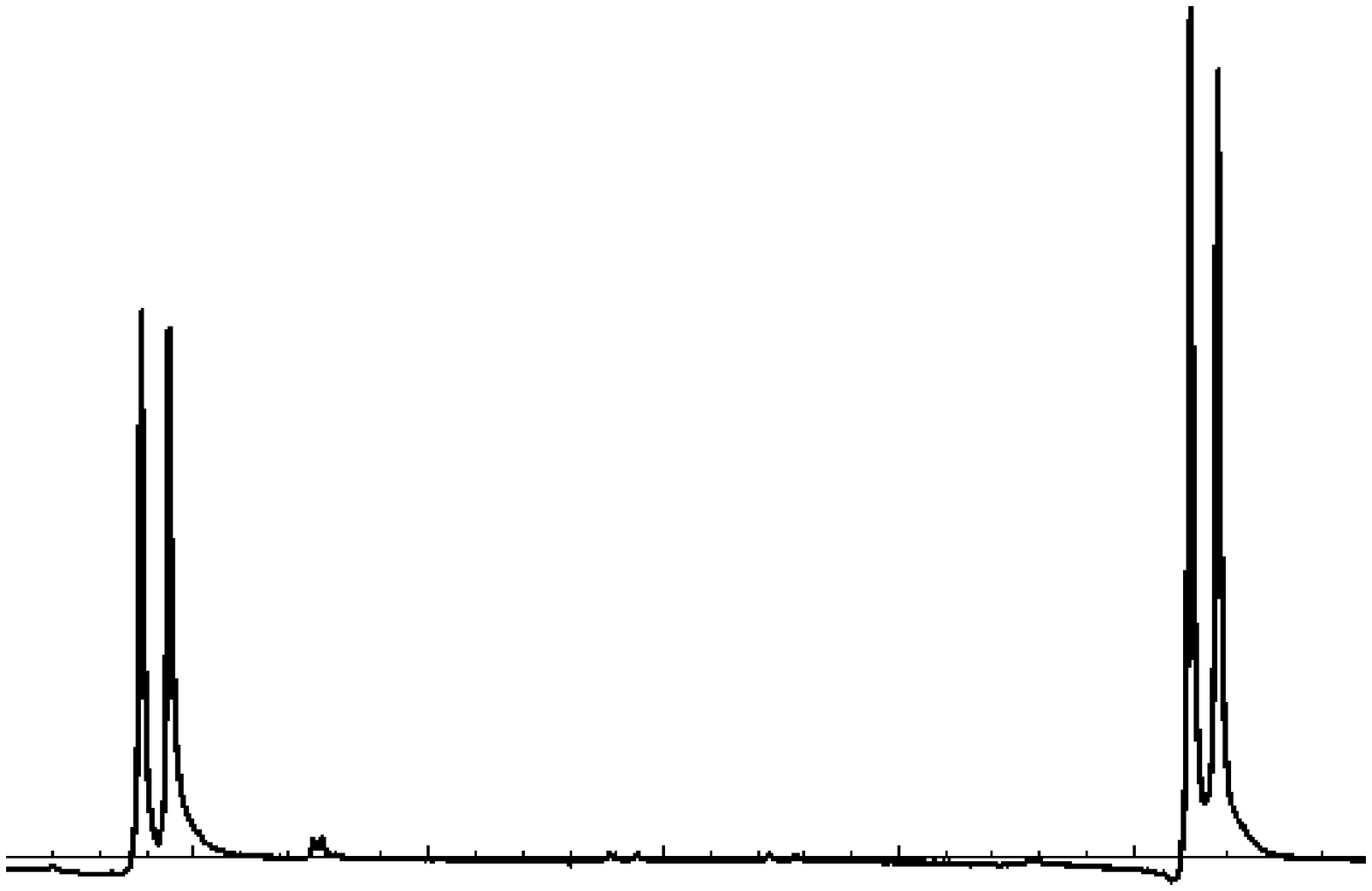}}
} \put(20,0){
\scalebox{0.66}[0.35]{\psfig{file=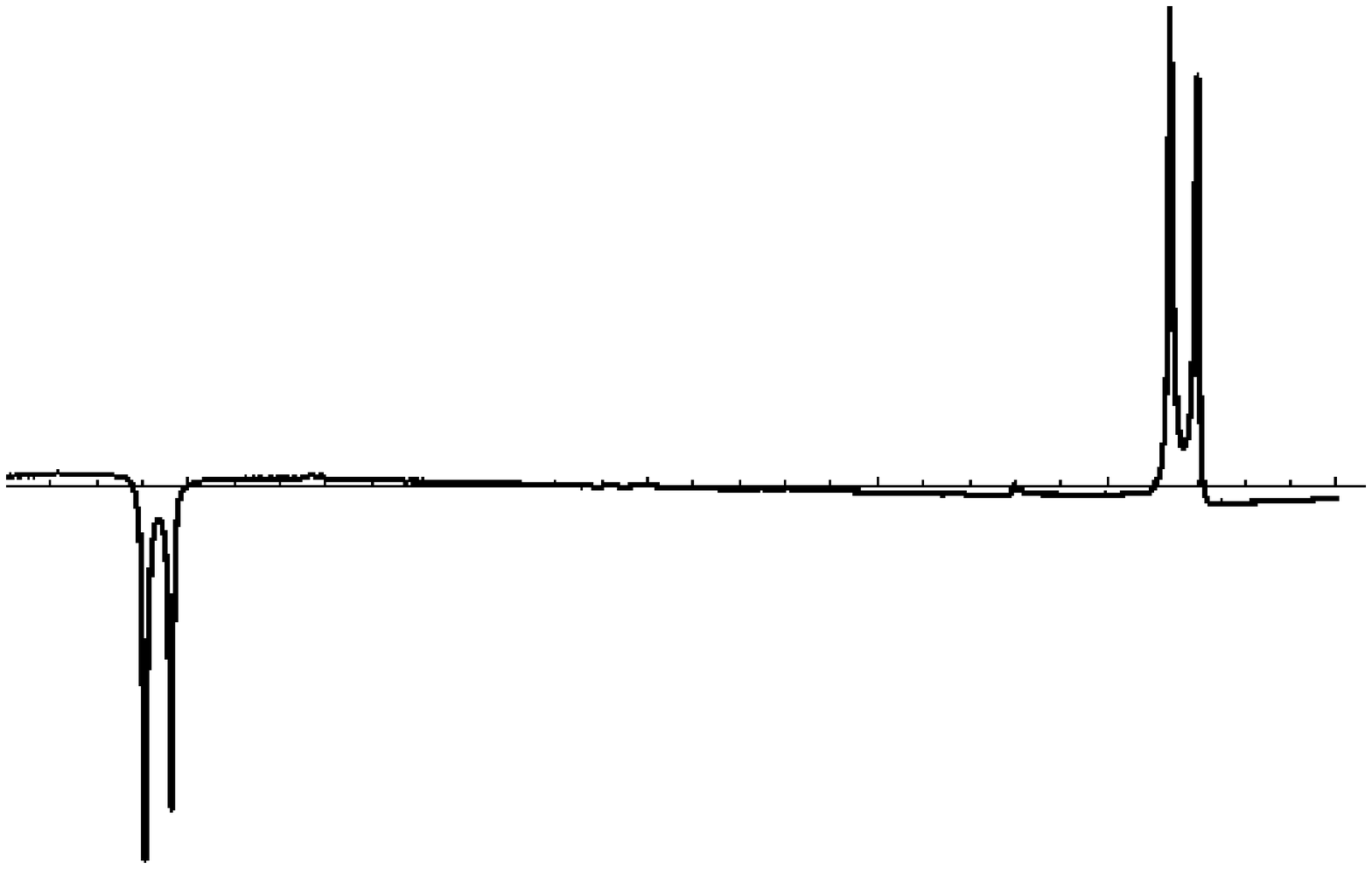}}
} \end{picture}
\caption{
The effect of applying the $[XOR\,]_{\mathrm{SC}}^1$
pulse sequence to the pseudo-pure state
$\mathbf I_z^1 + \mathbf I_z^2 - 2 \mathbf I_z^1 \mathbf I_z^2$
to get
$- \mathbf I_z^1 + \mathbf I_z^2 + 2 \mathbf I_z^1 \mathbf I_z^2$.
This is clearly seen from the change in sign of the first
in-phase doublet following a hard $[\pi/2]_y^{12}$ pulse,
before (above) and after (below) the XOR. }
\label{fig:pp2_scx_hard}
\end{figure}

The density matrices generated by these readout pulses can
be written in terms of the associated pseudo-spinors as
\begin{equation} \begin{aligned}
\half \mathbf 1 - (|0\rangle - |1\rangle) |1\rangle
(\langle 0| - \langle 1|) \langle 1|
& ~=~ \mathbf I_x^1 + \mathbf I_z^2
- 2\, \mathbf I_x^1 \mathbf I_z^2 \\
\half \mathbf 1 - |1\rangle (|0\rangle - |1\rangle)
\langle 1| (\langle 0| - \langle 1|)
& ~=~ \mathbf I_z^1 + \mathbf I_x^2
- 2\, \mathbf I_z^1 \mathbf I_x^2 ~.
\end{aligned} \end{equation}
In other words, each readout pulse puts the corresponding
spin into a superposition over its longitudinal states.
We can create a superposition over both spins by
performing the two readout pulses in rapid succession,
or with a single ``hard'' pulse (which takes only a small
fraction of the time required for a soft pulse), i.e.
\begin{equation}
\half \mathbf 1 - 2\, | 11 \rangle\langle 11 | \XRA{[\pi/2]_y^{12}\;}
\mathbf I_x^1 + \mathbf I_x^2 - 2\, \mathbf I_x^1 \mathbf I_x^2
\end{equation}
Since $2\,\mathbf I_x^1\mathbf I_x^2$ does not contribute to the signal,
the spectrum in this case consists of a pair of in-phase doublets,
exactly like that shown in Fig.~\ref{fig:eq2_hard}.

In NMR computing, superpositions over the basic
pseudo-pure states are always associated with nonzero
off-diagonal coherences in the density matrix.
The product operator above, for example, represents the matrix
\begin{equation} \begin{aligned}
\mathbf I_x^1 + \mathbf I_x^2 - 2\, \mathbf I_x^1 \mathbf I_x^2 ~=~
& \begin{aligned}[t] \half \bigl( \mathbf 1 \,-\,
& (| 0 \rangle - | 1 \rangle) (| 0 \rangle - | 1 \rangle) \\
& (\langle 0 | - \langle 1 |) (\langle 0 | - \langle 1 |) \bigr)
\end{aligned} \\
=~ & \half
\begin{pmatrix} ~0~ & ~1~ & ~1~ & ~-1~ \\ ~1~ & ~0~ & ~-1~ & ~1~ \\
~1~ & ~-1~ & ~0~ & ~1~ \\ ~-1~ & ~1~ & ~1~ & ~0~ \end{pmatrix}
\end{aligned} \end{equation}
Since the corresponding pseudo-spinor can be factored
into a product of one-spin pseudo-spinors (as shown),
it represents an ``unentangled'' state.

\begin{figure}[htb]
\begin{picture}(100,220)(15,0)
\put(20,115){
\scalebox{0.65}[0.30]{\psfig{file=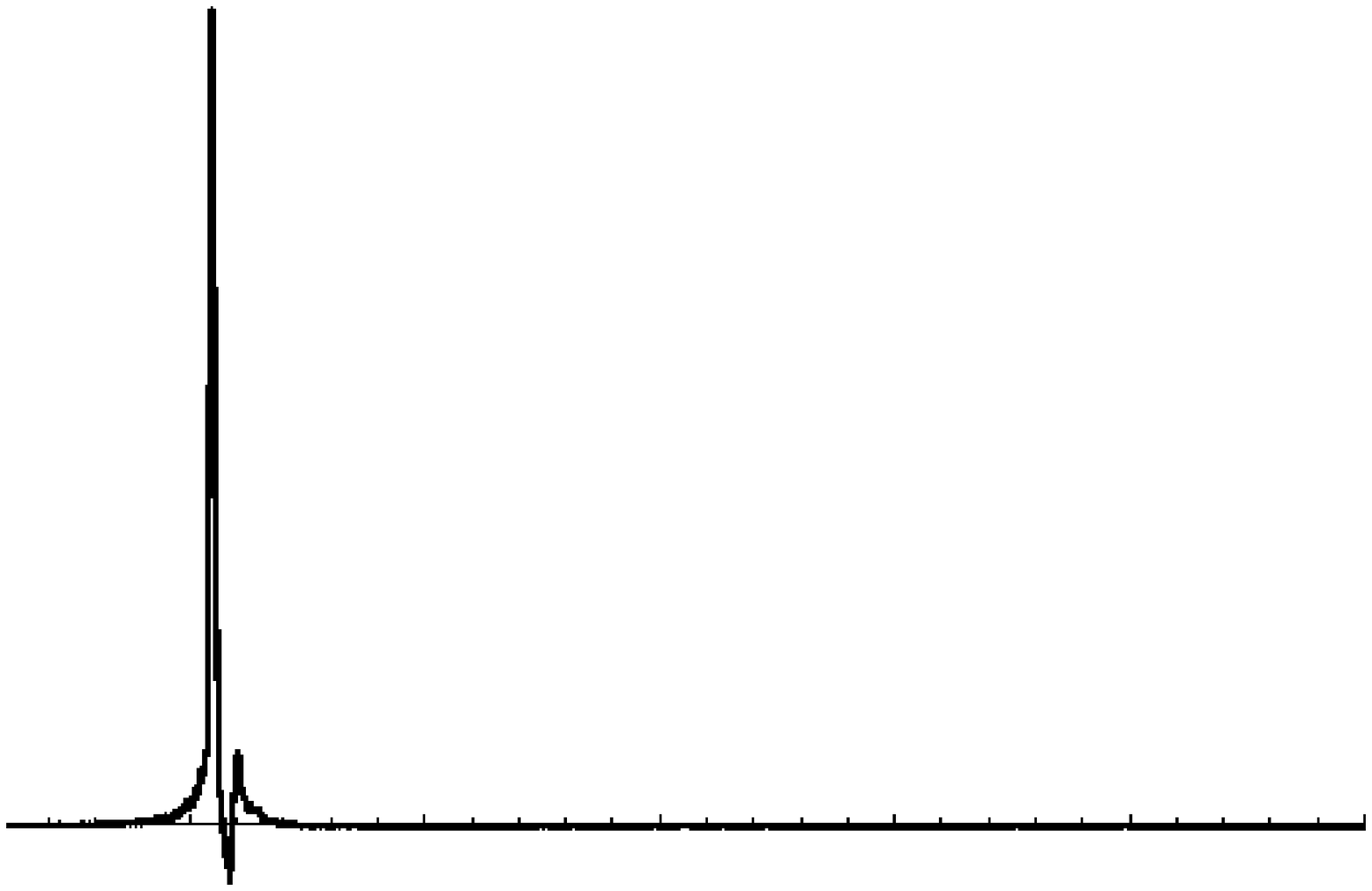}}
} \put(20,5){
\scalebox{0.65}[0.30]{\psfig{file=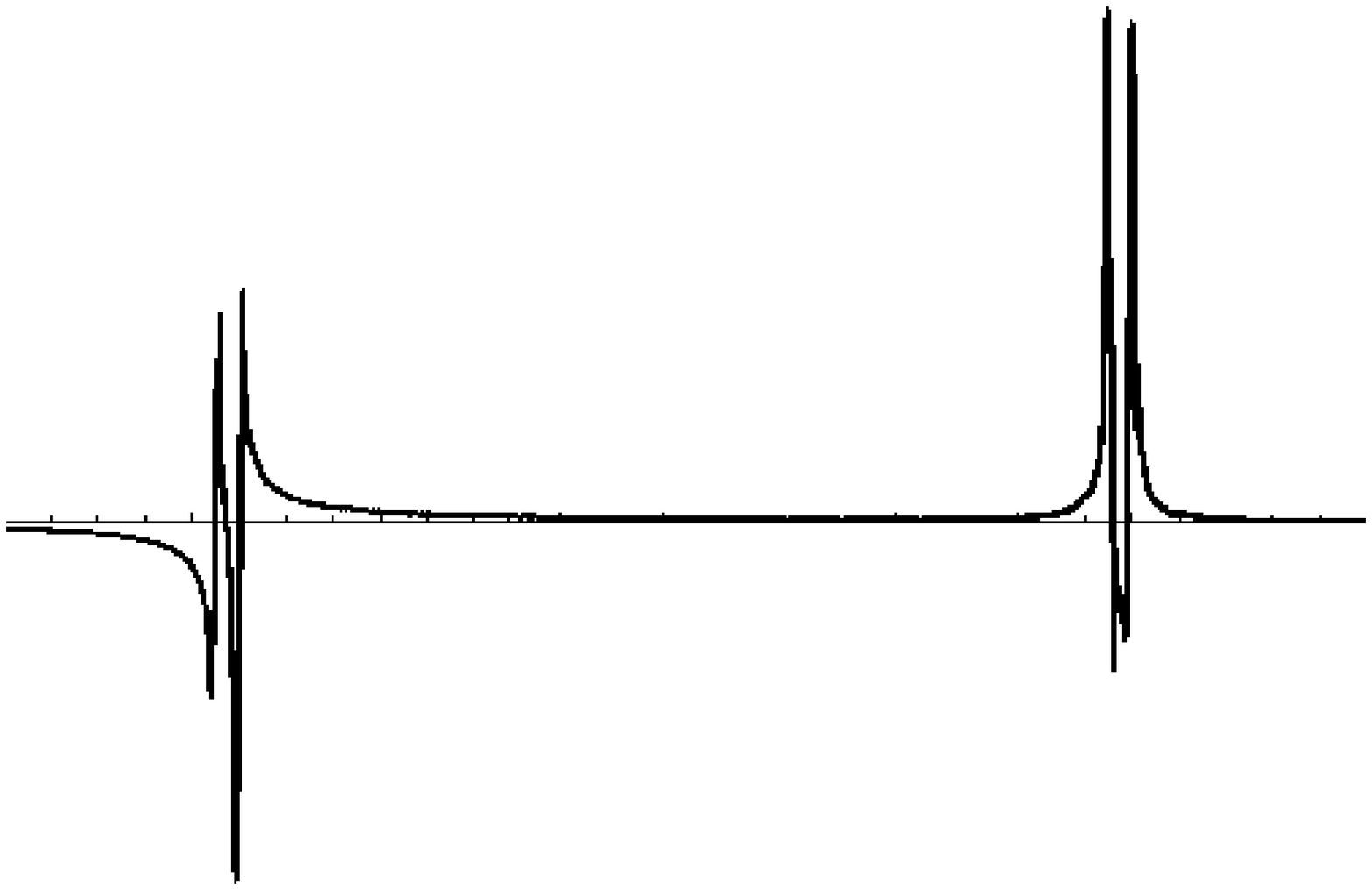}}
} \end{picture}
\caption{
The result of applying the $[XOR\,]_{\mathrm{SC}}^1$ sequence to the one-spin
superposition created by applying a $[\pi/2]_y^1$ pulse to the pseudo-pure
state $\mathbf I_z^1 + \mathbf I_z^2 + 2 \mathbf I_z^1 \mathbf I_z^2$,
to get $\mathbf I_y^1 + \mathbf I_z^2 + 2 \mathbf I_y^1 \mathbf I_z^2$ (above).
This is then further confirmed with an additional $[\pi/2]_y^2$ readout pulse
to get $\mathbf I_y^1 + \mathbf I_x^2 + 2 \mathbf I_y^1 \mathbf I_x^2$,
which consists of two in-phase doublets that
are $90^\circ$ out-of-phase with each other,
so that one doublet appears as a pair of dispersive peaks
whenever the other is phased to be an absorptive pair (below). }
\label{fig:pp2_super}
\end{figure}

The fact that we have successfully prepared a
pseudo-pure state, and our spin-coherence XOR gate,
can both be further corroborated by putting them together.
The resulting spectra are shown in Fig.~\ref{fig:pp2_scx_hard}.
We have also confirmed that the superposition
created by applying a soft $[\pi/2]_y^1$ pulse to the
$\mathbf I_z^1 + \mathbf I_z^2 + 2\,\mathbf I_z^2 \mathbf I_z^2$
state, namely $(|0\rangle + |1\rangle)|0\rangle$,
is changed by only inconsequential phase factors
on applying the $[XOR\,]_{\mathrm{SC}}^1$ sequence to it.
These phase changes can nevertheless be made visible by
applying a further $[\pi/2]_y^2$ readout pulse to the result,
as shown in Fig.~\ref{fig:pp2_super} and described in its caption.
Finally, we consider the result of applying the
$[XOR\,]_{\mathrm{SC}}^2$ gate, with its output on
the other spin, to this same one-spin superposition.
As is well-known, this creates an ``entangled'' state
\begin{equation}
\mathbf I_x^1 + \mathbf I_z^2 + 2\,\mathbf I_x^2 \mathbf I_z^2
~ \XRA{[XOR\,]_{\mathrm{SC}}^2\;} ~
2\, \mathbf I_x^1 \mathbf I_x^2 - 2\, \mathbf I_y^1 \mathbf I_y^2 +
2\,\mathbf I_z^2 \mathbf I_z^2 ~,
\end{equation}
or equivalently:
\begin{equation}
\begin{aligned}[c] \half \bigl( - & \mathbf 1 \,+\,
(| 00 \rangle + | 10 \rangle) \\ & (\langle 00 | + \langle 10 |)
\bigr) \end{aligned}
~ \XRA{[XOR\,]_{\mathrm{SC}}^2\;} ~
\begin{aligned}[c] \half \bigl( - & \mathbf 1 \,+\,
(| 00 \rangle + | 11 \rangle) \\ & (\langle 00 | + \langle 11 |)
\bigr) \end{aligned}
\label{eqn:entangled}
\end{equation}

\begin{figure}[tb]
\begin{picture}(300,100)(25,0)
\put(20,10) { \scalebox{0.30}[0.30]{\psfig{file=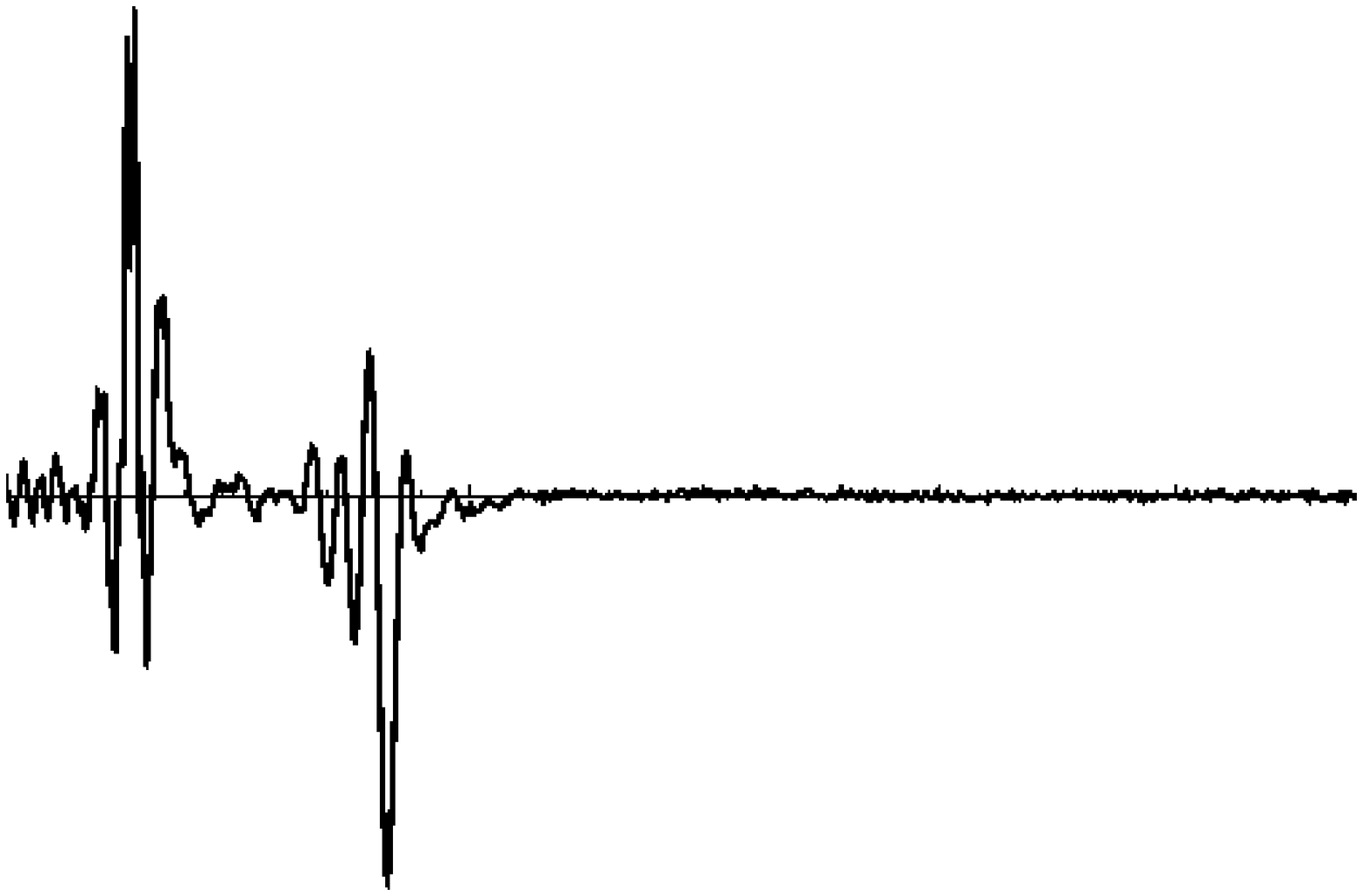}} }
\put(220,10) { \scalebox{0.30}[0.30]{\psfig{file=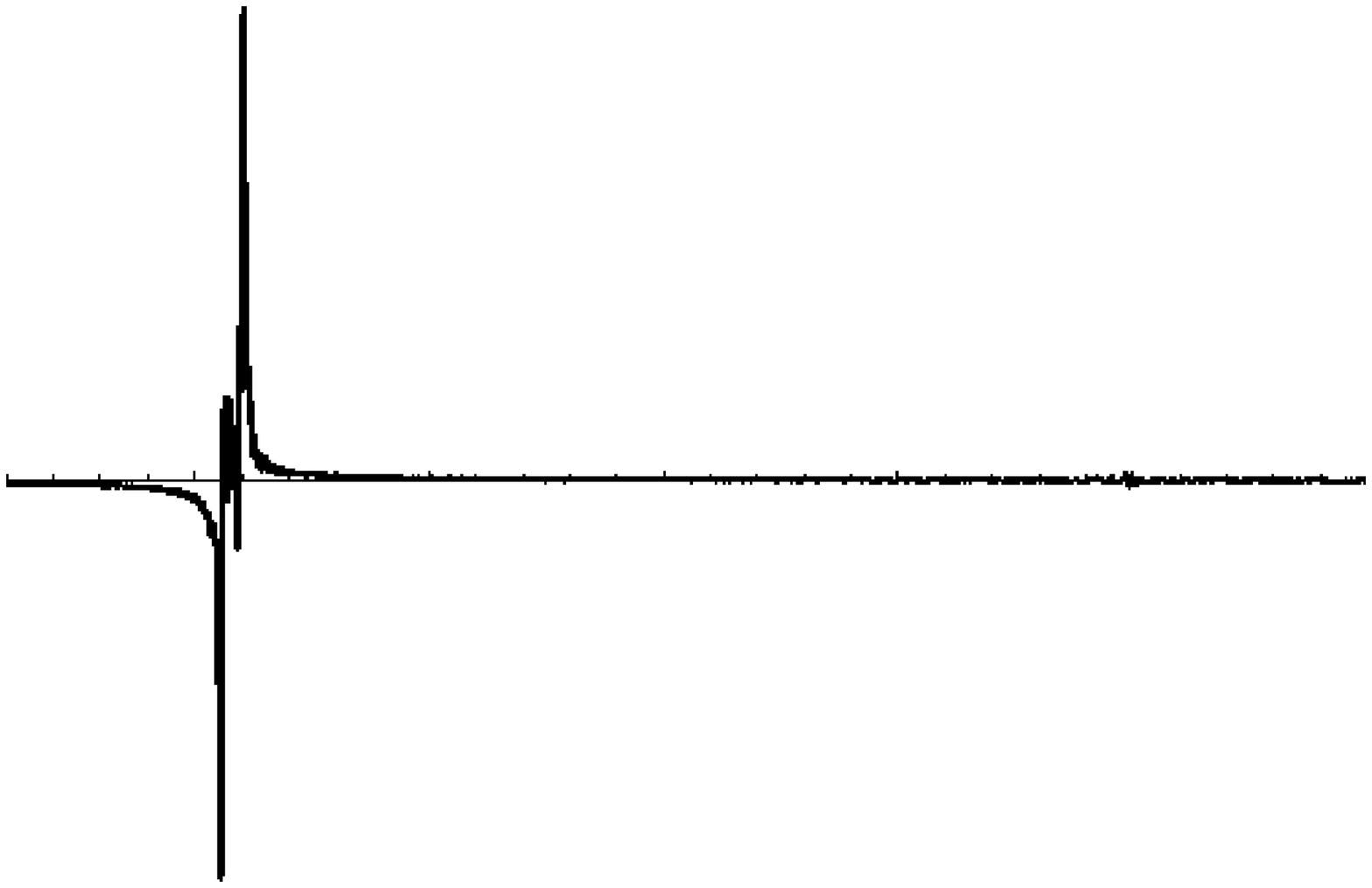}} }
\end{picture}
\caption{ Time-domain echo of the double-quantum coherence
characteristic of an ``entangled'' pseudo-spinor (left),
together with the corresponding frequency-domain spectrum (right).
The first echo, due to residual single-quantum coherence,
occurs in half the time required for the double-quantum
echo to form (see text). }
\label{fig:dqc_echo}
\end{figure}

In this case the transverse magnetization
corresponds to a \emph{double-quantum coherence},
\begin{equation}
2\,\mathbf I_x^1 \mathbf I_x^2 - 2\, \mathbf I_y^1 \mathbf I_y^2
~=~ \begin{pmatrix} ~0~&~0~&~0~&~1~\\ ~0~&~0~&~0~&~0~\\
~0~&~0~&~0~&~0~\\ ~1~&~0~&~0~&~0~ \end{pmatrix} ~,
\end{equation}
which precesses at twice the rate of the single-quantum
terms, but produces no observable magnetization.
The most direct way to confirm the creation
of this double-quantum coherence is
via a \emph{gradient echo} experiment.
In this technique, one applies a
$+z$-gradient for a period of $1/(2J_{12})$,
followed by a soft $[-\pi/2]_y^2$ pulse,
and then an \emph{inverse} $-z$-gradient.
The double-quantum coherence dephases at twice
the rate of a  single-quantum coherence,
but even though no macroscopic magnetization remains
at the end of the $+z$-gradient, the microscopic coherence
$2\, \mathbf I_x^1 \mathbf I_x^2$ is still converted to
$2\, \mathbf I_x^1 \mathbf I_z^2$ by the $[-\pi/2]_y^2$ pulse.
This term then rephases under the $-z$-gradient
at the usual rate for a single-quantum coherence,
while at the same time evolving under coupling to $\mathbf I_y^1$.
This in turn results in observable transverse magnetization,
which reaches its maximum after of period of $1/J_{12}$.
The double-quantum echo appears after twice the time of the echo
due to the rephasing of the (residual!) single-quantum coherence.
If we collect the data for a spectrum
starting right after the $[-\pi/2]_y^2$ pulse,
we observe the expected anti-phase doublet
(Fig.~\ref{fig:dqc_echo}).

For completeness, Table \ref{tab:pp3_coefs} gives
the coefficients of the seven diagonal product operators
in the eight basic pseudo-pure states of a three-spin system.
A pulse sequence and gradient sequence that
generates the $|000\rangle$ state is:
\begin{equation} \begin{aligned}
\, & \mathbf I_z^1 + \mathbf I_z^2 + \mathbf I_z^3 \\
\XRA[\hspace{5cm}]{[5\pi/12]_y^1-[\pi/3]_y^2-[\text{grad}]_z} ~&
\quart \mathbf I_z^1 + \half \mathbf I_z^2 + \mathbf I_z^3 \\
\XRA[\hspace{5cm}]{[\pi/4]_y^2-[1/(2J_{12})]-[-\pi/4]_x^2-[\text{grad}]_z} ~&
\quart \mathbf I_z^1 + \quart \mathbf I_z^2 + \mathbf I_z^3
+ \half \mathbf I_z^1 \mathbf I_z^2 \\
\XRA[\hspace{5cm}]{[\pi/4]_y^3 - [1/(2J)] - [\pi/4]_y^3 - [\text{grad}]_z} ~&
\begin{aligned}[t] & \quart \mathbf I_z^1 + \quart \mathbf I_z^2
+ \half \mathbf I_z^3 + \half \mathbf I_z^1 \mathbf I_z^2 + \\
& 2\, \mathbf I_z^1 \mathbf I_z^2 \mathbf I_z^3 \end{aligned} \\
\XRA[\hspace{5cm}]{[\pi/4]_y^3 - [1/(4J)] - [\pi/4]_x^3 - [\text{grad}]_z} ~&
\begin{aligned}[t] & \quart \mathbf I_z^1 + \quart \mathbf I_z^2
+ \quart \mathbf I_z^3 + \half \mathbf I_z^1 \mathbf I_z^2 + \\
& \half \mathbf I_z^1 \mathbf I_z^3 + \half \mathbf I_z^2 \mathbf I_z^3
+ \mathbf I_z^1 \mathbf I_z^2 \mathbf I_z^3 \end{aligned}
\end{aligned} \end{equation}
In this sequence, $[1/(2J_{12})]$ stands for a coupling
evolution period with the coupling constants $J_{13}$ and
$J_{23}$ averaged to zero by a $[\pi]$ pulse on the third spin,
while $[1/(2J)]$ and $[1/(4J)]$ are evolution
periods with $J_{12}$ averaged to zero and
$J_{13}$, $J_{23}$ averaged to the same value $J$.

\begin{table}[tb] \begin{center}
\renewcommand{\arraystretch}{1.2}
\begin{tabular}{|l||r|r|r|r|r|r|r|r|}
\hline
\raisebox{0pt}[14pt][8pt]{Spinor} &
$\mathbf I_z^1$ & $\mathbf I_z^2$ & $\mathbf I_z^3$ &
$2 \mathbf I_z^1 \mathbf I_z^2$ & $2 \mathbf I_z^1 \mathbf I_z^3$ & $2
\mathbf I_z^2 \mathbf I_z^3$ & $4 \mathbf I_z^1 \mathbf I_z^2 \mathbf I_z^3$ \\
\hline \hline
$|000\rangle$ & $ 1$ & $ 1$ & $ 1$ & $ 1$ & $ 1$ & $ 1$ & $ 1$ \\
$|001\rangle$ & $ 1$ & $ 1$ & $-1$ & $ 1$ & $-1$ & $-1$ & $-1$ \\
$|010\rangle$ & $ 1$ & $-1$ & $ 1$ & $-1$ & $ 1$ & $-1$ & $-1$ \\
$|011\rangle$ & $ 1$ & $-1$ & $-1$ & $-1$ & $-1$ & $ 1$ & $ 1$ \\
$|100\rangle$ & $-1$ & $ 1$ & $ 1$ & $-1$ & $-1$ & $ 1$ & $-1$ \\
$|101\rangle$ & $-1$ & $ 1$ & $-1$ & $-1$ & $ 1$ & $-1$ & $ 1$ \\
$|110\rangle$ & $-1$ & $-1$ & $ 1$ & $ 1$ & $-1$ & $-1$ & $ 1$ \\
$|111\rangle$ & $-1$ & $-1$ & $-1$ & $ 1$ & $ 1$ & $ 1$ & $-1$ \\
\hline
\end{tabular}
\caption{Table of coefficients of the seven diagonal product operators
corresponding to each basic pseudo-spinor for a three-spin system.}
\label{tab:pp3_coefs}
\end{center} \end{table}

\section{Conclusions}
We have demonstrated that nuclear magnetic
resonance spectroscopy provides an experimentally
accessible paradigm for quantum computing.
In particular, the results given in this paper
include the first physical implementations
of all the basic quantum logic gates,
including the XOR and Toffoli gates,
which have up to now been largely theoretical constructions.
We have further shown that one can actually
prepare a macroscopic ensemble of weakly
polarized spin systems in a pseudo-pure state,
which can be described by a spinor just like a true pure state,
and confirmed that one can put these states
into the equivalent of entangled superpositions.
Finally, we have shown how the state of the system
can be efficiently determined from spectra
collected following suitable readout pulses.

These operations provide essentially all the
ingredients needed to efficiently emulate a
true quantum computer by NMR spectroscopy.
Nevertheless, as has recently been
forcefully pointed out \cite{Warren:97},
the population difference for any single spin
that results from preparing a pseudo-pure state
by averaging over unitary transformations of
the thermal equilibrium state falls off as at
least $n/(2^n-1)$ with the number of spins $n$
(cf.\ \cite{Knill:97}).
This means that the signal-to-noise in the
spectra decreases exponentially with $n$,
which precludes extending this emulation
beyond $8 - 12$ spins in the foreseeable future.
Subject to these same limitations, however,
an NMR computer is also able to directly estimate the
expectation values of its observables \cite{CorFahHav:97}.
While this does not lead to any asymptotic performance
gains over what can be done with a quantum computer,
it remains a potentially significant advantage.
We conclude that the computational potential of
NMR spectroscopy, and of ensemble quantum computing
more generally, has yet to be fully explored.

\bigskip
\begin{center}
\textbf{Acknowledgements}\par\smallskip
This work was supported by NSF/DMR 9357603
to D.G.C., and by NSF/MCB 9527181 to T.F.H.
We thank Amr F.\ Fahmy for valuable discussions,
and Robert Griffin, Tommaso Toffoli and Gerhard
Wagner for their support of our efforts.
\end{center}

\bibliographystyle{plain}
\bibliography{../../chem,../../csci,../../nmr,../../self}

\end{document}